\documentclass[sigconf]{acmart}

\usepackage{booktabs}
\usepackage{xspace}
\usepackage{xcolor}
\usepackage{graphicx}
\usepackage{enumitem}
\usepackage{subcaption}
\usepackage{float}
\usepackage{natbib}
\usepackage{import}
\setlist[itemize]{noitemsep, topsep=0pt}
\setlist[enumerate]{noitemsep, topsep=0pt}
\newlist{requirements}{enumerate}{1}
\setlist[requirements, 1]{label = \textbf{R\arabic*:}}
\newlist{challenges}{enumerate}{1}
\setlist[challenges, 1]{label = \textbf{C\arabic*:}}

\graphicspath{ {./images/} }
\ifdefined\showcomments
\usepackage{todonotes}
\usepackage{letltxmacro}
\LetLtxMacro{\todonote}{\todo}
\renewcommand{\todo}[2][]
{\todonote[inline, caption={#2}, size=\footnotesize, #1]
{\renewcommand{\baselinestretch}{0.5}\selectfont#2\par}}
\else
\usepackage[disable]{todonotes}
\fi
\ifdefined\showchanges
\newcommand{\changed}[1]{\color{blue}{#1}\color{black}}
\else
\newcommand{\changed}[1]{#1}
\fi

\ccsdesc[500]{Security and privacy~Web protocol security}
\ccsdesc[300]{Security and privacy~Hardware-based security protocols}
\ccsdesc[300]{Security and privacy~Network security}
\ccsdesc[300]{Security and privacy~Privacy protections}
\setcopyright{rightsretained}
\copyrightyear{2019}
\acmYear{2019}
\acmConference[ACSAC '19]{2019 Annual Computer Security Applications Conference}{December 9--13, 2019}{San Juan, PR, USA}
\acmBooktitle{2019 Annual Computer Security Applications Conference (ACSAC '19), December 9--13, 2019, San Juan, PR, USA}
\acmPrice{15.00}
\acmDOI{10.1145/3359789.3359793}
\acmISBN{978-1-4503-7628-0/19/12}

\newcommand{\acron}{\textsf{\emph{PDoT}}\xspace}
\newcommand{\ANS}{\textsf{ANS}\xspace}
\newcommand{\NS}{\textsf{NS}\xspace}
\newcommand{\RA}{{\ensuremath{\sf{RA}}}\xspace}
\newcommand{\RR}{\textsf{\changed{RecRes}}\xspace}

\acmBadgeR[https://www.acsac.org/2019/]{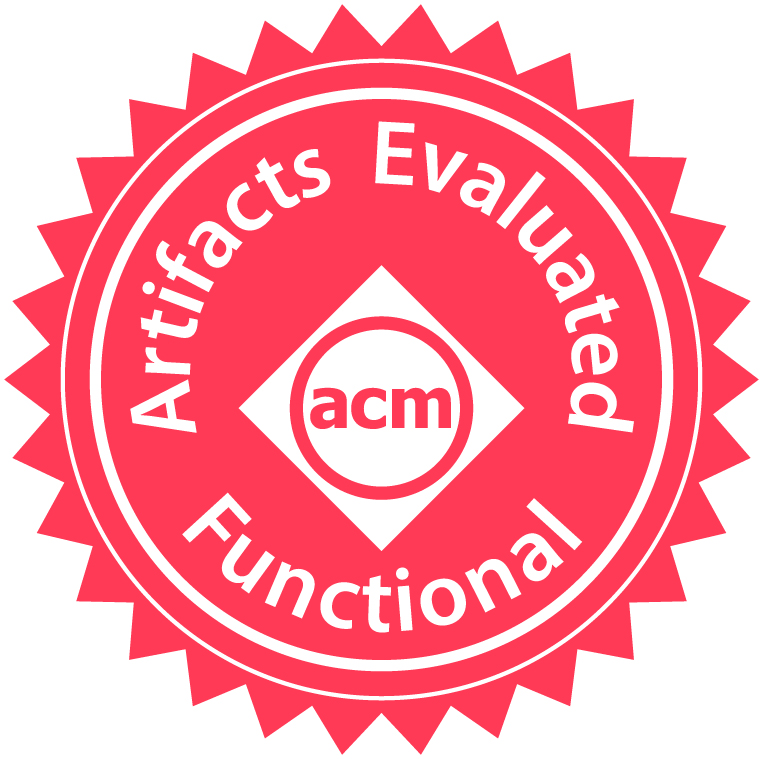}

\begin{document}
\title{{\tt\bf PDoT}: Private DNS-over-TLS with TEE Support}

\author{Yoshimichi Nakatsuka}
\affiliation{
  \institution{University of California, Irvine}
}
\email{nakatsuy@uci.edu}

\author{Andrew Paverd}
\authornote{Work done while visiting University of California, Irvine, as a US-UK Fulbright Cyber Security Scholar.}
\affiliation{
  \institution{Microsoft Research}
}
\email{andrew.paverd@ieee.org}

\author{Gene Tsudik}
\affiliation{
  \institution{University of California, Irvine}
}
\email{gene.tsudik@uci.edu}

\begin{abstract}
Security and privacy of the Internet Domain Name System (DNS) have been longstanding concerns.
Recently, there is a trend to protect DNS traffic using Transport Layer Security (TLS).
However, at least two major issues remain: (1) how do clients authenticate DNS-over-TLS 
endpoints in a scalable and extensible manner; and (2) how can clients trust endpoints to 
behave as expected? In this paper, we propose a novel Private DNS-over-TLS (\acron) architecture.
\acron includes a DNS Recursive Resolver (\RR) that operates within a Trusted Execution Environment (TEE). 
Using \emph{Remote Attestation}, DNS clients can authenticate, and receive strong assurance of trustworthiness of \acron \RR. 
We provide an open-source proof-of-concept implementation of \acron and use it to experimentally demonstrate that its 
latency and throughput match that of the popular Unbound DNS-over-TLS resolver.
\end{abstract}

\keywords{Domain Name System, Privacy, Trusted Execution Environment}

\maketitle

\section{Introduction}\label{sec:intro}
The Domain Name System (DNS)~\cite{DNS} is a distributed system that translates human-readable domain names into IP addresses.
It has been deployed since 1983 and, throughout the years, DNS privacy has been a major concern.

In 2015, Zhu et al.~\cite{DoTOakland} proposed a DNS design that runs over Transport Layer Security (TLS) connections~\cite{TLS}.
DNS-over-TLS protects privacy of DNS queries and prevents man-in-the-middle (MiTM) attacks against DNS responses. 
\cite{DoTOakland} also demonstrated the practicality of DNS-over-TLS in real-life applications.
Several open-source recursive resolver (\RR) implementations, including Unbound~\cite{Unbound} and Knot 
Resolver~\cite{KnotResolver}, currently support DNS-over-TLS.
In addition, commercial support for DNS-over-TLS has been increasing, e.g., Android P devices~\cite{AndroidP} and Cloudflare's \texttt{1.1.1.1} \RR~\cite{Cloudflare}.
However, despite attracting interest in both academia and industry, some problems remain.

The first challenge is how clients authenticate the \RR. Certificate-based authentication is natural for websites, 
since the user (client) knows the URL of the desired website and the certificate securely binds this URL to a public key.
However, the same approach cannot be used to authenticate a DNS \RR because the \RR does not have a URL or any other unique long-term user-recognizable identity that can be included in the certificate.
One way to address this issue is to provide clients with a white-list of trusted \RR-s' public keys.
However, this is neither scalable nor maintainable, because the white-list would have to include all 
possible \RR operators, ranging from large public services (e.g., \texttt{1.1.1.1}) to small-scale providers, e.g., a local \RR provided by a coffee-shop.

Even if the \RR can be authenticated, the second major issue is the lack of means to determine whether a given \RR is trustworthy.
For example, even if communication between client stub (client) and \RR, and between \RR and the 
name server (\NS) is authenticated and encrypted using TLS, the \RR must decrypt the DNS query in order to resolve it and contact the relevant \NS-s. 
\changed{This allows the \RR to learn unencrypted DNS queries, which poses privacy risks of a malicious \RR misusing 
the data, e.g., profiling users or selling their DNS data. Some \RR operators go to great lengths to assure users that their data is private.
For example, Cloudflare promises \emph{``We will never sell your data or use it to target ads''} and goes on to say \emph{``We've retained 
KPMG to audit our systems annually to ensure that we're doing what we say''}~\cite{Cloudflare}.
Although helpful, this still requires users to trust the auditor and can only be used by operators who can afford an auditor.}

In this paper, we use Trusted Execution Environments (TEEs) and \emph{Remote Attestation} (\RA) to address these two problems.
By using \RA, the identity of the \RR is no longer relevant, since clients can check what software a given \RR is running and make trust decisions based on how the \RR behaves.
\RA is one of the main features of modern hardware-based TEEs, such as Intel Software 
Guard Extensions (SGX)~\cite{SGX} and ARM TrustZone~\cite{TZ}.  Such TEEs are now widely available, with Intel CPUs after the 
7th generation supporting SGX, and ARM Cortex-A CPUs supporting TrustZone. 
TEEs with \RA capability are also available in cloud services, such as Microsoft Azure~\cite{AzureSGX}.
Specifically, our contributions are:
\begin{itemize}
  \item We design a Private DNS-over-TLS (\acron) architecture, the main component of which is
  a privacy-preserving \RR that operates within a commodity TEE. Running the \RR inside a TEE 
  prevents even the \RR operator from learning clients' DNS queries, thus providing query privacy.
  Our \RR design addresses the authentication challenge by enabling clients to trust the 
  \RR based on how it behaves, and not on who it claims to be. (See Section~\ref{sec:sys_model_challenges}).
  \item We implement a proof-of-concept \acron \RR using Intel SGX and evaluate its security, deployability, and performance.
  All source code and evaluation scripts are publicly available~\cite{PDoTcode}.
  Our results show that \acron handles DNS queries without leaking information while achieving 
  sufficiently low latency and offering acceptable throughput (See Sections~\ref{sec:impl}~and~\ref{sec:eval}).
  \item In order to quantify privacy leakage via traffic analysis, we performed an Internet measurement study. It 
  shows that 94.7\% of the top $1,000,000$ domain names can be served from a \emph{privacy-preserving}
   \NS that serves at least two distinct domain names, and 65.7\% from a \NS that serves 100+ domain 
   names. (See Section~\ref{sec:discussion}).
\end{itemize}

\section{Background}\label{sec:background}
\subsection{Domain Name System (DNS)}\label{sec:dns}
DNS is a distributed system that translates host and domain names into IP addresses. 
DNS includes three types of entities: \textit{Client Stub} (client), \textit{Recursive Resolver} (\RR), and 
\textit{Name Server} (NS). Client runs on end-hosts. It receives DNS queries from applications, 
creates DNS request packets, and sends them to the configured \RR.  Upon receiving a request, 
\RR sends DNS queries to NS-s to resolve the query on client's behalf. 
When NS receives a DNS query, it responds to \RR with either the DNS \emph{record} that answers 
client's query, or the IP address of the next NS to contact. \RR thus recursively queries NS-s until the 
record is found or a threshold is reached.  The NS that holds the queried record is called: 
\textit{Authoritative Name Server} (ANS). After receiving the record from ANS, \RR forwards it to 
client. It is common for \RR to cache records so that repeated queries can be handled more efficiently.
\subsection{Trusted Execution Environment (TEE)}\label{sec:tee}
A Trusted Execution Environment (TEE) is a security primitive that isolates code and data from privileged 
software such as the OS, hypervisor, and BIOS. All software running outside TEE is considered untrusted.
Only code running within TEE can access data within TEE, thus protecting confidentiality and integrity 
of this data against untrusted software. Another typical TEE feature is remote attestation (\RA), which allows 
remote clients to check precisely what software is running inside TEE.

One recent TEE example is Intel SGX, which enables applications to create isolated execution environments 
called \textit{enclaves}. The CPU enforces that only code running within an enclave can access that enclave's data.
SGX also provides \RA functionality. 

\textbf{Memory Security.} SGX reserves a portion of memory called the Enclave Page Cache (EPC). 
It holds 4KB pages of code and data associated with specific enclaves. 
EPC is protected by the CPU to prevent non-enclave access to this memory region. 
Execution threads enter and exit enclaves using SGX CPU instructions, thus ensuring that in-enclave code execution can only begin from well-defined call gates.  
From a software perspective, untrusted code can make \textit{ECALLs} to invoke enclave functions, and enclave code can make \textit{OCALLs} to invoke untrusted functions outside the enclave.

\textbf{Attestation Service.} SGX provides two types of attestation: local and remote. 
Local attestation enables one enclave to attest another (running on the same machine)
to verify that the latter is a genuine enclave actually running on the same CPU. 
Remote attestation involves more entities. 
First, an application enclave to be attested creates a \textit{report} that summarizes 
information about itself, e.g., code it is running.  This report is sent to a special enclave,
called \textit{quoting enclave} which is provided by Intel and available on all SGX machines. 
Quoting enclave confirms that requesting application enclave is running on the same machine 
and returns a \textit{quote}, which is a report with the quoting enclave's signature. 
The application enclave sends this quote to the Intel Attestation Service (IAS) and obtains an 
\textit{attestation verification report}. This is signed by the IAS confirming that the application enclave 
is indeed a genuine SGX enclave running the code it claims.  Upon receiving an attestation verification report, 
the verifier can make an informed trust decision about the behavior of the attested enclave.

\textbf{Side-Channel Attacks.} SGX is vulnerable to side-channel 
attacks~\cite{CacheSideChannel, ControllChannelSideChannel}, and various mechanisms 
have been proposed~\cite{Sanctum, TSGX, Tamrakar2017} to mitigate them.
Since defending against side-channel attacks is orthogonal to our work,
we expect that a production implementation would include relevant mitigation mechanisms.

\section{Adversary Model \& Requirements}\label{sec:adv_model_sys_req}
\subsection{Adversary Model}\label{sec:adv_model}
The adversary's goal is to learn, or infer, information about DNS queries sent by clients.
We consider two types of adversaries, based on their capabilities:

The first type is a malicious \RR operator who has full control over the physical machine, 
its OS and all applications, including \RR. We assume that the adversary cannot break any cryptographic 
primitives, assuming that they are correctly implemented. We also assume that it cannot physically attack 
hardware components, e.g., probe CPU physically to learn TEE secrets. This adversary also controls all of 
\RR's communication interfaces, allowing it to drop/delay packets, measure the time required for query 
processing, and observe all cleartext packet headers.
The second type is a network adversary, which is strictly weaker than the malicious \RR operator.
\changed{In the passive case, this adversary can observe any packets that flow into and out of \RR.
In the active case, this adversary can modify and forge network packets.
DNS-over-TLS alone (without \acron) is sufficient to thwart a passive network adversary.
However, since an active adversary could redirect clients to a malicious \RR, clients need an
efficient mechanism to authenticate the \RR and determine whether it is trustworthy, 
which is the main contribution of \acron.}

We do not consider Denial-of-Service (DoS) attacks on \RR, since these do not help to achieve either 
adversary's goal of learning clients' DNS queries. Connection-oriented \RR-s can defend against DoS 
attacks using cookie-based mechanisms to prevent SYN flooding~\cite{DoTOakland}.

\subsection{System Requirements}\label{sec:sys_req}
We define the following requirements for the overall system:
\begin{requirements}
  \item \textbf{Query Privacy.} Contents of client's query (specifically, domain name to be resolved) should 
  not be learned by the adversary. Ideally, payload of the DNS packets should be encrypted.
  However, even if packets are encrypted, their headers leak information, 
  such as source and destination IP addresses. In Section~\ref{sec:ppans}, we quantify the amount of 
  information that can be learned via traffic analysis.
  \item \textbf{Deployability.} Clients using a privacy-preserving \RR should require no special hardware. 
  Minimal software modifications should be imposed. Also, for the purpose of transition and compatibility,
  a privacy-preserving \RR should be able to interact with legacy clients that only support unmodified DNS-over-TLS.
  \item \textbf{Response Latency.} A privacy-preserving \RR should achieve similar response latency to that of a regular \RR.
  \item \textbf{Scalability.} A privacy-preserving \RR should process a realistic volume of queries 
  generated by a realistic number of clients. 
\end{requirements}

\changed{
\noindent {\em Note:} query privacy guarantees provided by \acron rely on the forward-looking assumption 
that communication between \RR and respective NS-s is also protected by DNS-over-TLS.
The DNS Privacy (DPrive) Working Group is working towards a standard for encryption and authentication of 
DNS resolver-to-ANS communication~\cite{dns-res-to-auth}, using essentially the same mechanism as 
DNS-over-TLS. We expect an increasing number of NS-s to begin supporting this standard in the near future.
Once \acron is enabled at the \RR, it can provide incremental query privacy for queries served from a DNS-over-TLS NS.
As discussed in Section~\ref{sec:impl}, with small design modifications, \acron can be adapted for use in NS-s. 
}

\section{System Model \& Design Challenges}\label{sec:sys_model_challenges}
\subsection{\acron System Model}\label{sec:sys_model}
\begin{figure}[!t]
  \centering
  \includegraphics[width=\columnwidth]{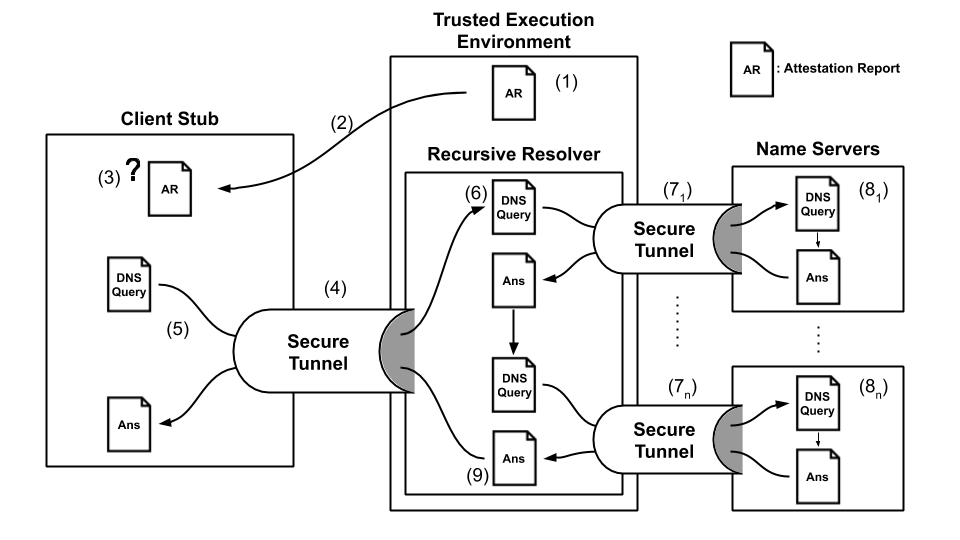}
  \caption{Overview of the proposed system.}
  \Description{Overview of the proposed system.}
  \label{fig:system_model}
\end{figure}

Figure~\ref{fig:system_model} shows an overview of \acron. It includes four types of entities: client, \RR, TEE, NS-s. 
We now summarize \acron operation, reflected in the figure:
(1) After initial start-up, TEE creates an attestation report.
(2) When client initiates a secure TLS connection, the attestation report is sent from \RR to the client 
alongside all other information required to setup a secure connection.
(3) Client authenticates and attests \RR by verifying the attestation report. It checks whether \RR 
is running inside a genuine TEE and running trusted code.
(4) Client proceeds with the rest of the TLS handshake procedure only if verification succeeds. 
(5) Client sends a DNS query to \RR through the secure TLS channel it has just set up.
(6) \RR receives a DNS query from client, decrypts it into TEE memory, and learns the domain name that the 
client wants to resolve.
(7) \RR sets up a secure TLS channel to the appropriate NS in order to resolve the query.
(8) \RR sends a DNS query to NS over that channel. If NS's reply includes an IP address of the
next NS, \RR sets up another TLS channel to that NS. This is done repeatedly, until \RR successfully 
resolves the name to an IP address.
(9) Once \RR obtains the final answer, it sends this to client over the secure channel. Client 
can reuse the TLS channel for future queries.

\changed{
Note that we assume \RR is not under the control of the user.
In some cases, users could run their own \RR-s, which would side-step the concerns about query privacy.
For example, modern home routers are sufficiently powerful to run an in-house \RR.
However, this approach cannot be used in public networks (e.g., airports or coffee shop WiFi networks), which are the target scenarios for \acron. 
}

\subsection{Design Challenges}\label{sec:challenges}
The following key challenges were encountered in the process of \acron's design:
\begin{challenges}
  \item \textbf{TEE Limited Functionality.} In order to satisfy their security requirements, TEE-s often limit 
  the functionality available to code that runs within them. One example is the inability to fork within the 
  TEE. Forking a process running inside the TEE forces the child process to run outside the TEE, breaking 
  \RR security guarantees. Another example is that system calls, such as socket communication, 
  cannot be made from within the TEE.
  \item \textbf{TEE Memory Limitations.} A typical TEE has a relatively small amount of memory. 
  Although an SGX enclave can theoretically have a large amount of in-enclave memory, this will require page swapping of EPC pages. 
  The pages to be swapped must be encrypted and integrity protected in order to meet the security requirements of SGX. Therefore, page swapping places a heavy burden on performance. To avoid page swapping, enclave size 
  should be less than the size of the EPC -- typically, 128MB. Since \RR is a performance-critical 
  application, its size should ideally not exceed 128MB. This limit negatively impacts \RR throughput, 
  as it bounds the number of threads that can be spawned in a TEE.
  \item \textbf{TEE Call-in/Call-out Overhead.} Applications requiring functionality that is not available 
  within the TEE must switch to the non-TEE side. 
  This introduces additional overhead, both from the switching itself, and from the need to flush and reload CPU caches. 
  Identifying and minimizing the number 
  of times \RR switches back and forth (whilst keeping \RR functionality correct) is a substantial challenge.
\end{challenges}

\begin{figure*}[!t]
  \centering
  \includegraphics[width=0.70\textwidth]{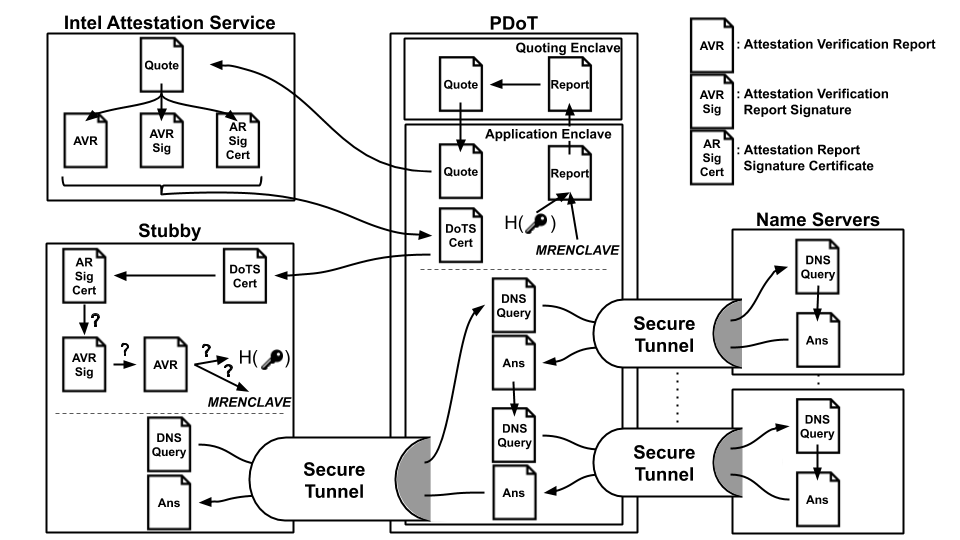}
  \caption{Overview of \acron implementation.}
  \Description{Overview of \acron implementation.}
  \label{fig:implementation_detail}
\end{figure*}

\section{Implementation}\label{sec:impl}
Figure~\ref{fig:implementation_detail} shows an overview of the \acron design.
\changed{
Since our design is architecture-independent, it can be implemented on any TEE architecture that 
provides the features outlined in Section~\ref{sec:tee}.  We chose the off-the-shelf Intel SGX as the platform 
for the proof-of-concept \acron implementation in order to conduct an accurate performance evaluation on 
real hardware. (See Section~\ref{sec:eval}). Therefore, our implementation is subject to performance and memory 
constraints in the current version of Intel SGX, and is thus best suited for small-scale networks,  e.g., the public WiFi 
network provided of a typical coffee shop. However, as TEE technology advances, we expect that our design will 
scale to larger networks.
}

\subsection{\acron}\label{sec:dots}
\acron consists of two parts: (1) trusted part residing in TEE enclaves, and (2) untrusted part that 
operates in the non-TEE region. The former is responsible for resolving DNS queries, and the latter --
for accepting incoming connections, assigning file descriptors to sockets, and sending/receiving data 
received from the trusted part.

\textbf{Enclave Startup Process.} When the application enclave starts, it generates a new public-private 
key-pair within the enclave. It then creates a \textit{report} that summarizes enclave and platform state. 
The report includes a SHA256 hash of the entire code that is supposed to run in the enclave (called
\textit{MRENCLAVE} value) and other attributes of the target enclave. \acron also includes a SHA256 
hash of the previously generated public key in the report. The report is then passed on to the SGX quoting 
enclave to receive a \textit{quote}. The quoting enclave signs the report
and thus generates a quote, which cryptographically binds the public key to the application enclave. 
The quoting enclave sends the quote to the application enclave, which forwards it to the Intel Attestation 
Service (IAS) to obtain an \textit{attestation verification report}. It can be used in the future by clients 
to verify the link between the public key and the MRENCLAVE value. After receiving the attestation verification 
report from IAS, the application enclave prepares a self-signed X.509 certificate required for the TLS 
handshake. In addition to the public key, the certificate includes: (1) attestation verification report, 
(2) attestation verification report signature, and (3) attestation report signing certificate,  extracted from (1). 
MRENCLAVE value and hash of public key are enclosed in the attestation verification report. 

\textbf{TLS Handshake Process.}\footnote{In implementing this process, we heavily relied on SGX RA
TLS~\cite{SGXRATLS} whitepaper.} Once the application enclave is created, \acron can 
create TLS connections and accept DNS queries from clients. The client initiates a TLS handshake 
process by sending a message to \acron. This message is captured by untrusted part of \acron and 
triggers the following events.\footnote{Since we consider a malicious \RR operator, it has an option not to 
trigger these events. However, clients will notice that their queries are not being answered and 
can switch to a different \RR.} First, untrusted part of \acron tells the application enclave to create a 
new TLS object within the enclave for this incoming connection. This forces the TLS endpoint to reside 
inside the enclave. The TLS object is then connected to the socket where the client is waiting to be served. 
\RR then exchanges several messages with the client, including the self-signed certificate that was created in the 
previous section. Having received the certificate from \RR, the client authenticates \RR and validates the 
certificate. (For more detail, see Section~\ref{sec:client}). Only if the authentication and validation succeed, 
the client resumes the handshake process.

\textbf{DNS Query Resolving Process.} The client sends a DNS query over the TLS channel
established above. Upon receiving the query, \RR decrypts it within the application enclave 
and obtains the target domain name. \RR begins to resolve the name starting from root NS, by doing the following 
repeatedly: 1) set up a TLS channel with NS, 2) send DNS queries and receive replies via that channel. 
Once \RR receives the answer from NS, \RR returns it to the client over the original TLS channel.

\begin{figure}[t!]
  \centering
  \includegraphics[width=\columnwidth]{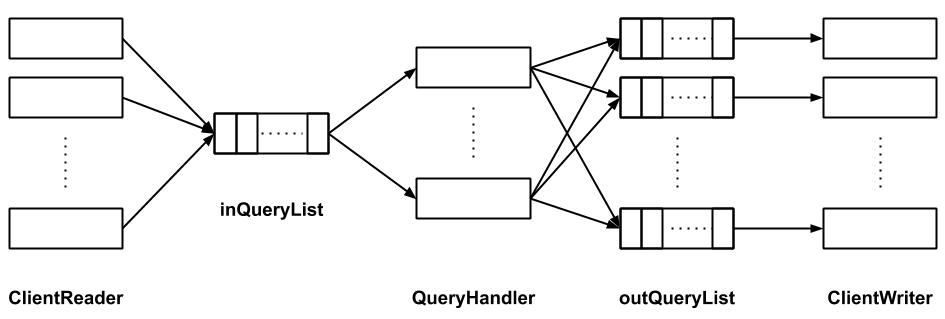}
  \caption{Overview of \acron threading model.}
  \Description{Overview of \acron threading model.}
  \label{fig:threading_model}
\end{figure}

Figure~\ref{fig:threading_model} illustrates how \acron divides DNS query resolution process into three 
threads: (1) receiving DNS query -- \textit{ClientReader}, (2) resolving it -- \textit{QueryHandler}, and 
(3) returning the answer -- \textit{ClientWriter}.

ClientReader and ClientWriter threads are spawned anew upon each query. Dividing receiving and sending 
processes and giving them a dedicated thread is helpful because many clients send multiple DNS queries
within a short timespan without waiting for the answer to the previous query.\footnote{For example, a client has 
received a webpage that includes images and advertisements that are served from servers located at different 
domains. This triggers multiple DNS queries at the same time.} When ClientReader thread receives a DNS 
query from the client, it stores the query and a client ID in a FIFO queue, called \textit{inQueryList}.

QueryHandler threads are spawned when \acron starts up. The number of QueryHandler threads is configured 
by \RR operator. QueryHandler threads are shared among all current ClientReader and ClientWriter threads. 
When a QueryHandler thread detects an entry in the inQueryList, it removes this entry and retrieves the query 
and the client ID. QueryHandler first checks whether this client is still accepting answers from \RR. If not,  
QueryHandler simply ignores this query and moves on to the next one. If the client is still accepting answers, 
QueryHandler resolves the query and puts the answer into a FIFO queue (called \textit{outQueryList}) 
dedicated to that specific client.

In some cases, NS response might be too slow. When that happens, QueryHandler thread gives up on 
resolving that particular query and moves on to the next query, since it is very likely that the request was 
dropped. This also prevents resources (such as mutex) from being locked up by this QueryHandler thread. 
In our implementation, this timeout was set to be the same as the client's timeout, since there is no point in sending 
the answer to the client after that.

Once an answer is added to outQueryList dedicated to its client, ClientWriter uses that answer to compose
a DNS reply packet and sends it to the client. The reason we have $N$ outQueryLists for $N$ clients 
is to improve performance. With only one outQueryList, ClientWriter threads must search through the 
queue to find the answer for the connected client. This takes $O(M\times~N)$ time, where $N$ is the 
number of clients and $M$ is the number of queries each client sends. Instead, with $N$ 
outQueryLists, we reduce complexity to $O(1)$ because ClientWriter thread merely selects the query 
at the head of the list.

\changed{

\textbf{Caching.}
Some DNS recursive resolvers can cache query results. Caching is beneficial from the client's perspective,
since, in case of a cache hit, the \RR can answer immediately, thus reducing query latency.
The \RR also benefits from not having to establish connections to external \NS-s.
However, irrespective of how it is implemented, caching at the \RR causes potential privacy leaks, e.g., 
timings can reveal whether a certain domain record was already in the cache. This is an orthogonal challenge, discussed 
in Section~\ref{sec:caching}.

To explore caching in a privacy-preserving resolver, we implemented a simple in-enclave cache for \acron.
It uses a red-black tree data structure and stores all records associated with the clients' queries, indexed by the 
queried domain. This results in $O(log_2(N))$ access times with $N$ entries in the cache.
In practice, \acron could also use current techniques to mitigate against side-channel attacks on 
cache's memory access patterns, e.g., \cite{sasy2017zerotrace, costa2017pyramid, Tamrakar2017}.
During remote attestation, clients can ascertain whether the resolver has enabled caching, and which mitigations it uses.

\textbf{\acron \ANS with TEE support.} With minor design changes, \acron \RR design can be modified for use as an \ANS.
Similar to the caching mechanism described above, an \acron \ANS can look up the answers to queries in an internal database, 
rather than contact external \NS-s. The same way that clients authenticate \acron \RR, the \RR can authenticate the \acron \ANS.
Clients can thus establish trust in both \RR and \ANS using \emph{transitive attestation}~\cite{alder2019sfaas}.

}

\subsection{Client with \acron Support}\label{sec:client}
We took the Stubby client stub from the {\tt getdns} project~\cite{Stubby} which offers DNS-over-TLS 
support and modified it to perform remote attestation during the TLS handshake. 
We now describe how the client verifies its \RR, decides whether the \RR is trusted, and 
emits the DNS request packet.

\textbf{\RR Verification.} After receiving a DNS request from an application, the client first checks 
whether there is an existing TLS connection to its \RR. If so, the client reuses it. If not, it 
attempts to establish a new connection. During the handshake, the client receives a certificate 
from \RR, from which it extracts: 1) attestation verification report, 2) attestation verification report signature, 
and 3) attestation report signing certificate. This certificate is self-signed by IAS and we assume that the 
client trusts it. From (3), the client first retrieves the IAS public key and, using it, verifies (2). Then, the client 
extracts the SHA256 hash of \RR's public key from (1) and verifies it against a copy from (3). This way,
the client is assured that \RR is indeed running in a genuine SGX enclave and uses this public key for the 
TLS connection.

\textbf{Trust Decision.} The client also extracts the MRENCLAVE value from (1), which it compares 
against the list of acceptable MRENCLAVE values. If the MRENCLAVE value is not listed or one of the 
verification steps fail, the client stub aborts the handshake, moves on to the next \RR, and re-starts the process. 
Note that the trust decision process is different from the normal TLS trust decision process. Normally, a TLS 
server-side certificate binds the public key to one or more URLs and organization names. However, by 
binding the MRENCLAVE value with the public key, the clients can trust \RR based on its behavior, and 
not its organization (recall that the MRENCLAVE value is a hash of \RR code).
\changed{
There several options for deciding which MRENCLAVE values are trustworthy.
For example, vendors could publish lists of expected MRENCLAVE values for their resolvers.
For open-source resolvers like \acron, anyone can re-compute the expected MRENCLAVE value by 
recompiling the software, assuming a reproducible build process.
This would allow trusted third parties (e.g., auditors) to inspect the source code, ascertain that it upholds 
required privacy guarantees, and publish their own lists of trusted MRENCLAVE values. 
}

\textbf{Sending DNS request.} Once the TLS connection is established, the client sends the DNS query 
to \RR over the TLS tunnel. If it does not receive a response from \RR within the specified timeout, it 
assumes that there is a problem with \RR and sends a DNS reply message to the application 
with an error code SERVFAIL.

\subsection{Overcoming Technical Challenges}\label{sec:overcoming_challenges}
As discussed in Section~\ref{sec:challenges}, \acron faced three main challenges, which we addressed as follows:

\textbf{Limited TEE Functionality.} 
The inability to use sockets within the TEE is a challenge because the \RR cannot communicate with the outside world. 
We address this issue by having a process running outside the TEE, as described in Section~\ref{sec:dots}. 
This process forwards packets from the client to TEE through ECALLs and sends packets received from TEE via OCALLs.
However, this processes might redirect the packet to a malicious process or simply drop it.
We discuss this issue in Section~\ref{sec:sec_analysis}.
Another function unavailable within TEE is forking a process. 
\acron uses \texttt{pthreads} instead of forking to run multiple tasks concurrently in a TEE.

\textbf{Limited TEE Memory.}
We use several techniques to address this challenge. 
First, we ensure no other enclaves (other than the quoting enclave) run on \RR machine. 
This allows \acron to use all available EPC memory.
Second, we fix the number of QueryHandler threads in order to save space. 
This is possible because of dis-association of QueryHandler and ClientReader/Writer threads.

\textbf{OCALL and ECALL Overhead.} 
ECALLs and OCALLs introduce overhead and therefore should be avoided as much as 
possible. For example, all  threads  mentioned in the previous section must wait until they receive the 
following information: for ClientReader thread -- DNS query from the client, for QueryProcessor thread -- 
query from inQueryList, and for ClientWriter thread -- response from outQueryList. \acron was implemented 
so that these threads wait inside the enclave. If we were to wait outside the enclave, we would have to use 
an ECALL to enter the enclave each time the thread proceeds.

\section{Evaluation}\label{sec:eval}
\subsection{Security Analysis}\label{sec:sec_analysis}
This section describes how query privacy (Requirement R1) is achieved, 
with respect to the two types of adversaries, per Section~\ref{sec:adv_model}.

\textbf{Malicious \RR operator.} 
Recall that a malicious \RR operator controls the machine that runs \acron \RR. It cannot obtain the query 
from intercepted packets since they flow over the encrypted TLS channel. Also, because the local
TLS endpoint resides inside the \RR enclave, the malicious operator cannot retrieve the query from the enclave,
as it does not have access to the protected memory region.

However, a malicious \RR operator 
may attempt to connect the socket to a malicious TLS server that resides in either: 1) an untrusted region, 
or 2) a separate enclave that the operator itself created. If the operator can trick the client into establishing a 
TLS connection with the malicious TLS server, the adversary can obtain the plaintext DNS queries. 
For case (1), the verification step at the client side fails because the TLS server certificate does
not include any attestation information. For case (2), the malicious enclave might receive a legitimate 
attestation verification report, attestation verification report signature, and attestation report signing certificate 
from IAS. However, that report would contain a different MRENCLAVE value, which would be rejected by the 
client. To convince the client to establish a connection with \acron \RR, the adversary has no choice except 
to run the code of \acron \RR. Therefore, in both cases, the adversary cannot trick the client into establishing a 
TLS connection with a TLS server other than the one running a \acron \RR.

\textbf{Network Adversary.} 
Recall that this adversary captures all packets to/from \acron. It cannot obtain the plaintext queries 
since they flow over the TLS tunnel. The only information it can obtain from packets includes
cleartext header fields, such as source and destination IP addresses. This information, coupled with a 
timing attack, might let the adversary correlate a packet sent from the client with a packet sent to an NS. 
The consequent amount of privacy leakage is discussed in Section~\ref{sec:ppans}

\subsection{Deployability}\label{sec:deployability}
Section~\ref{sec:impl} shows how \acron clients do not need special hardware, and require only minor software 
modifications (Requirement R2). To aid deployability, \acron also provides several configurable parameters, 
including: the number of QueryHandle threads (to adjust throughput), the amount of memory dedicated to 
each thread (to serve clients that send a lot of queries at a given time), and the timeout of QueryHandle threads 
(to adjust the time for a QueryHandle thread to acquire a resource).
Another consideration is incremental deployment, where some clients may request DNS-over-TLS without 
supporting \acron. \acron can handle this situation by having its TLS certificate {\bf also} signed by a trusted 
root CA, since legacy clients will ignore \acron-specific attestation information.

\changed{
On the client side, an ideal deployment scenario would be for browser or OS vendors to update their client stubs 
to support \acron{}.
The same way that browser vendors currently include and maintain a list of trusted root CA certificates in their browsers, 
they could include and periodically update a list of trustworthy MRENCLAVE values for PDoT resolvers.
This could all be done transparently to end users.
As with root CA certificates, expert users can manually add/remove trusted MRENCLAVE values for their own systems.
In practice, there are only a handful of recursive resolver software implementations. Thus, even allowing for multiple 
versions of each, the list of trusted MRENCLAVE values would be orders of magnitude smaller than the list of 
public keys of every trusted resolver, as would be required for standard DNS-over-TLS.
}

\begin{figure*}[t!]
  \centering
  \begin{subfigure}{0.49\textwidth}
    \centering
    \includegraphics[width=\textwidth]{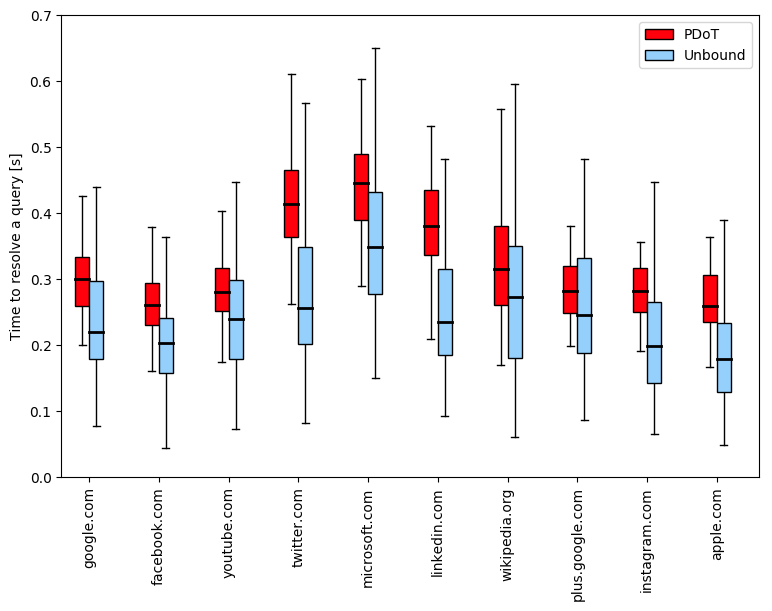}
    \caption{Latency of \acron and Unbound (Cold Start)}
    \label{fig:latency_eval_cold}
  \end{subfigure}
  \begin{subfigure}{0.49\textwidth}
    \centering
    \includegraphics[width=\textwidth]{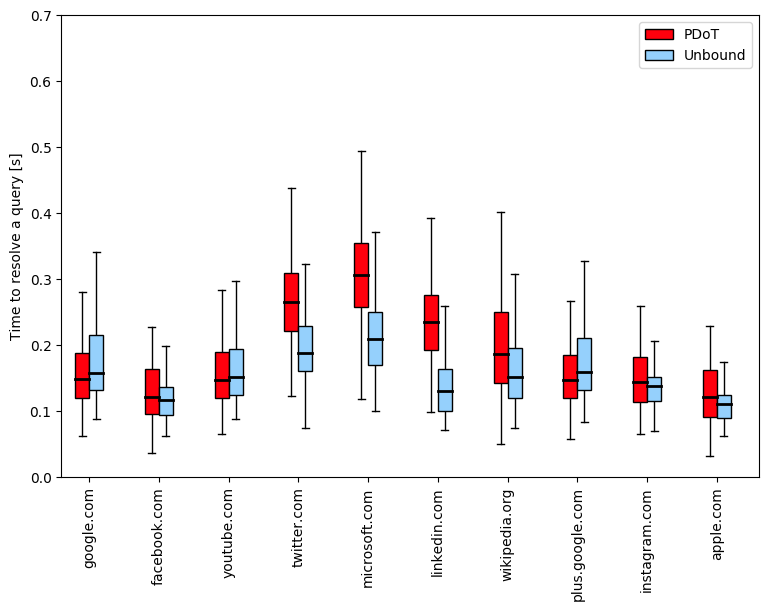}
    \caption{Latency of \acron and Unbound (Warm Start)}
    \label{fig:latency_eval_warm}
  \end{subfigure}
  \caption{Latency comparison of \acron and Unbound}
  \Description{Latency comparison of \acron and Unbound}
  \label{fig:latency_eval}
\end{figure*}

\subsection{Performance Evaluation}\label{sec:performance_eval}
We ran \acron on a low-cost Intel NUC consisting of an Intel Pentium Silver J5005 CPU with 128~MB of EPC 
memory and 4~GB of RAM. We used Ubuntu 16.04 and the Intel SGX SDK version~2.2.
We configured our \RR to support up to 50 concurrent clients and process queries using 30 QueryHandle threads.
For comparison, we performed the same benchmarks using Unbound~\cite{Unbound}, a popular open source \RR.

\subsubsection{Latency Evaluation}\label{sec:latency_eval}
The objective of our latency evaluation is to assess overhead introduced by running \RR inside an enclave. 
To do so, we measure the time to resolve a DNS query using \acron and compare with Unbound.
To meet requirement R3, \acron should not incur a significant increase in latency compared to Unbound.

\textbf{Experimental Setup.} The client and \RR ran on the same physical machine to remove network delay. 
We conducted the experiment using \acron and Unbound as the \RR, and Stubby as the client. We measured 
latency under two different scenarios: cold start and warm start. In the former, the client sets up a 
new TLS connection every time it sends a query to the \RR. In the warm start scenario, the client sets up one 
TLS connection with the \RR at the beginning, and reuses it throughout the experiment. In other words, 
the cold start measurements also include the time required to establish the TLS connection.
In this experiment, the caching mechanisms of both \acron and Unbound were disabled.

We created a python program to feed DNS queries to the client. The program sends 100 queries 
sequentially for ten different domains. That is, the program waits for an answer to the 
previous query before sending the next query.
We used the top ten domains of the Majestic Million domain list~\cite{MajesticMillion}. 

The python program measures the time between sending the query and receiving an answer. 
For the cold start experiment, we spawned a new Stubby client and established a new TLS connection for each query. 
In the warm start scenario, we first established the TLS connection by sending a query for another domain 
(not in the top ten), but did not include this in the timing measurement.

\changed{Note that the numeric latency values are specific to our experimental setup because they depend on 
network bandwidth of our \RR, and latency between the latter and relevant \NS-s.
The important aspect of this experiment is the ratio between the latencies of \acron and Unbound.
Therefore it is not meaningful to compute average latency over a large set of domains.
Instead, we took multiple measurements for each of a small set of domains (e.g., 100 measurements 
for each of 10 domains) so as to analyse the range of response latencies for each domain.}

\changed{
\textbf{Results and observations.} Results of latency measurements are are shown in Figure~\ref{fig:latency_eval}.
Red boxes show latency of \acron and the blue boxes -- of Unbound.
In these plots, boxes span from the lower to upper quartile values of collected data.
Whiskers span from the highest datum within the 1.5 interquartile range (IQR) of the upper 
quartile to the lowest datum within the 1.5 IQR of the lower quartile.
Median values are shown as black horizontal lines inside the boxes.

For the cold-start case in Figure~\ref{fig:latency_eval_cold}, although Unbound is typically faster than our 
proof-of-concept \acron implementation, the range of latencies is similar.
For 7 out of 10 domains, the upper whisker of \acron was lower than that of Unbound.
Overall, \acron shows an average 22\% overhead compared to Unbound in the cold-start setting.

For the warm-start case in Figure~\ref{fig:latency_eval_warm}, the median latency is lower across the 
board compared to the cold-start setting because 
the TLS tunnel has already been established.
In this setting, \acron shows an average of 9\% overhead compared to Unbound.
In practice, once the client has established a connection to \RR, it will maintain this connection; thus, 
the vast majority of queries will see only the warm-start latency.
}

\begin{figure*}[t!]
  \centering
  \begin{subfigure}{0.32\textwidth}
    \centering
    \includegraphics[width=\textwidth]{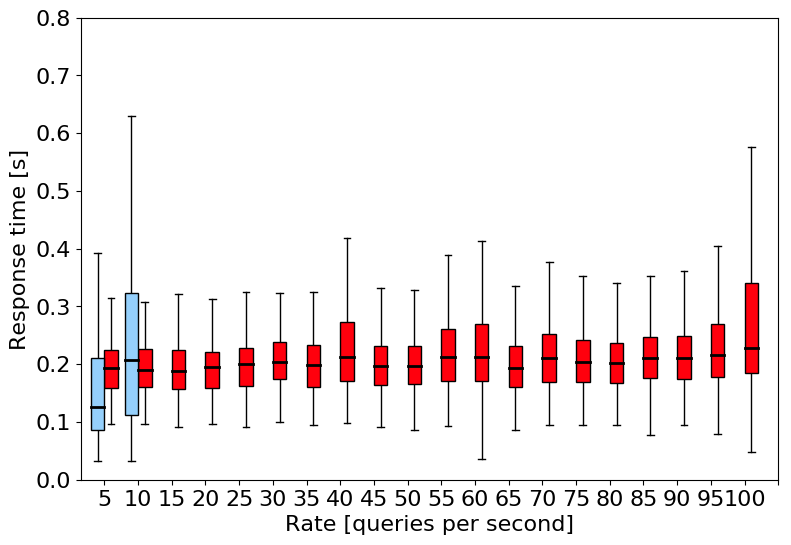}
    \caption{Throughput for 1 client}
    \label{fig:throughput_1}
  \end{subfigure}
  \begin{subfigure}{0.32\textwidth}
    \centering
    \includegraphics[width=\textwidth]{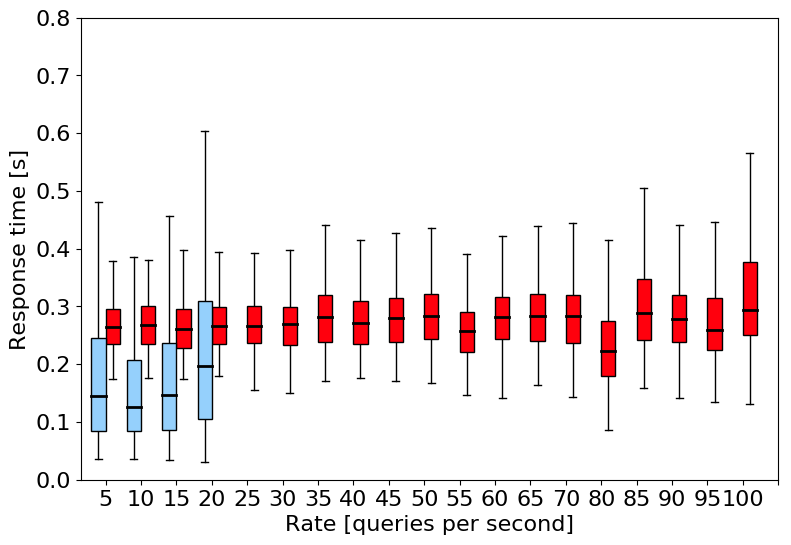}
    \caption{Throughput for 2 clients}
    \label{fig:throughput_2}
  \end{subfigure}
  \begin{subfigure}{0.32\textwidth}
    \centering
    \includegraphics[width=\textwidth]{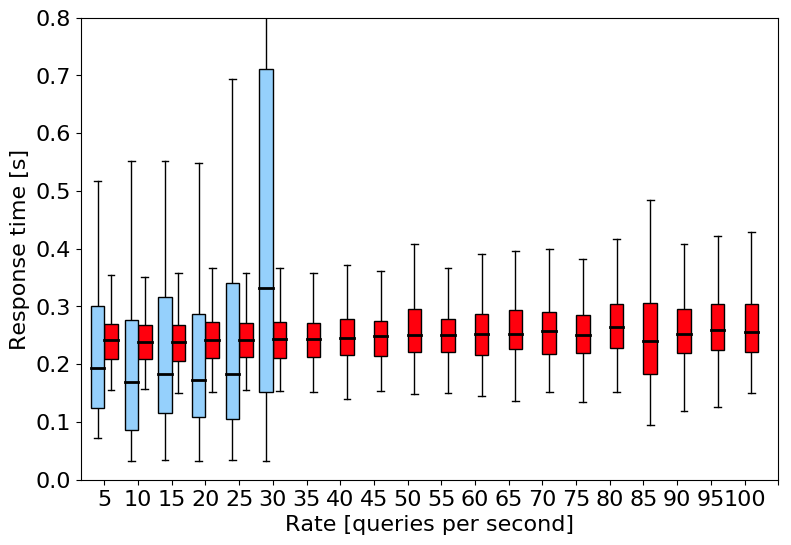}
    \caption{Throughput for 3 clients}
    \label{fig:throughput_3}
  \end{subfigure}
  \begin{subfigure}{0.32\textwidth}
    \centering
    \includegraphics[width=\textwidth]{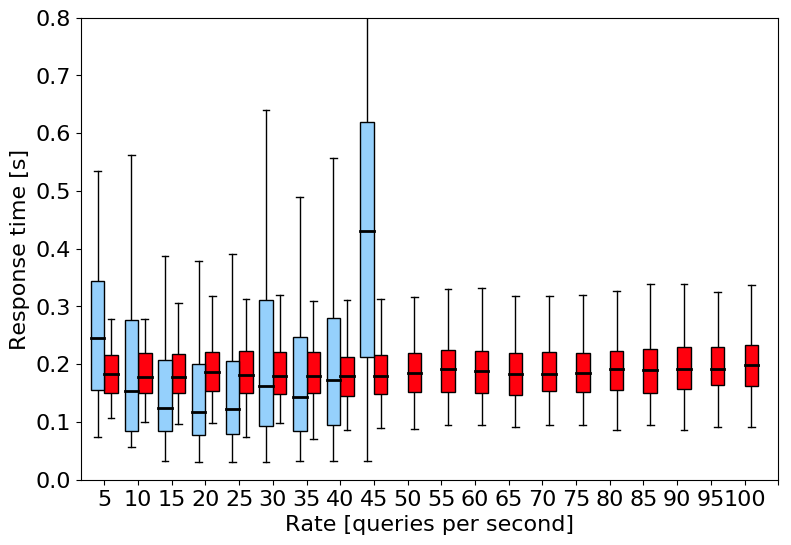}
    \caption{Throughput for 4 clients}
    \label{fig:throughput_4}
  \end{subfigure}
  \begin{subfigure}{0.32\textwidth}
    \centering
    \includegraphics[width=\textwidth]{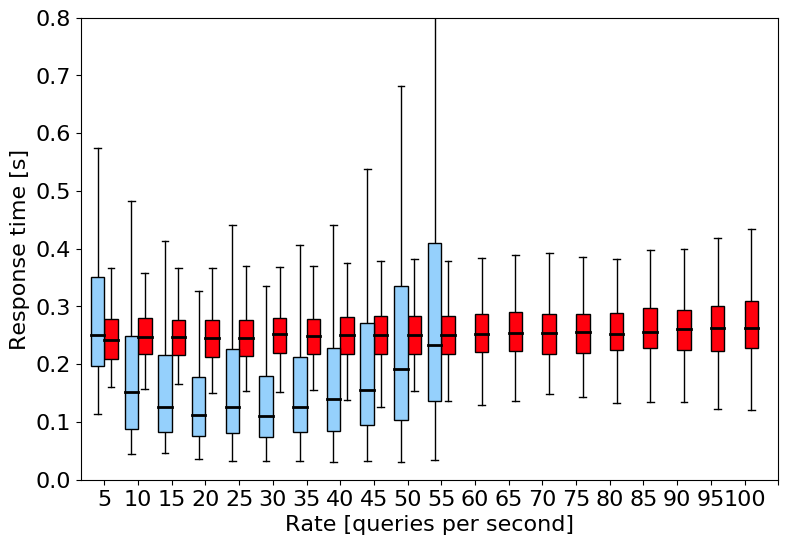}
    \caption{Throughput for 5 clients}
    \label{fig:throughput_5}
  \end{subfigure}
  \begin{subfigure}{0.32\textwidth}
    \centering
    \includegraphics[width=\textwidth]{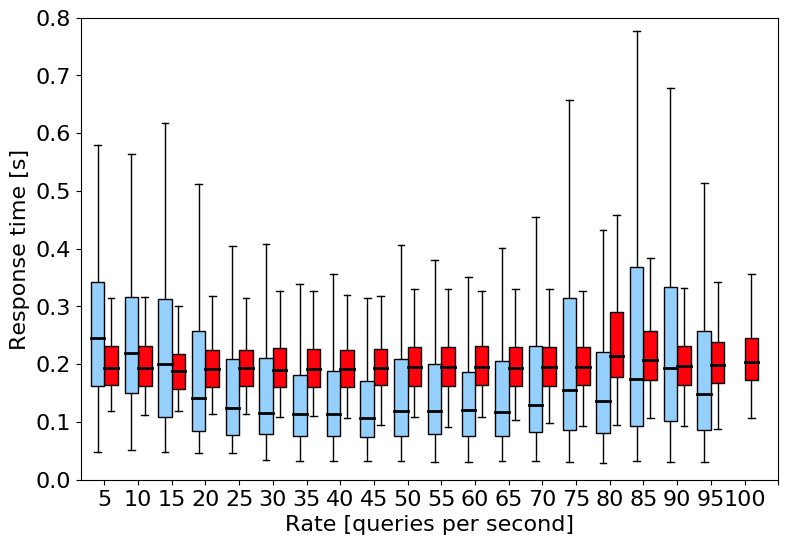}
    \caption{Throughput for 10 clients}
    \label{fig:throughput_10}
  \end{subfigure}
  \begin{subfigure}{0.32\textwidth}
    \centering
    \includegraphics[width=\textwidth]{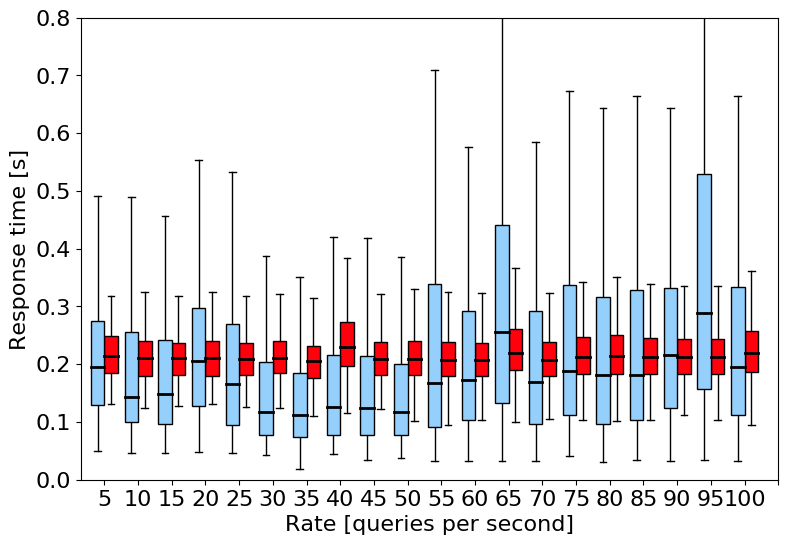}
    \caption{Throughput for 15 clients}
    \label{fig:throughput_15}
  \end{subfigure}
  \begin{subfigure}{0.32\textwidth}
    \centering
    \includegraphics[width=\textwidth]{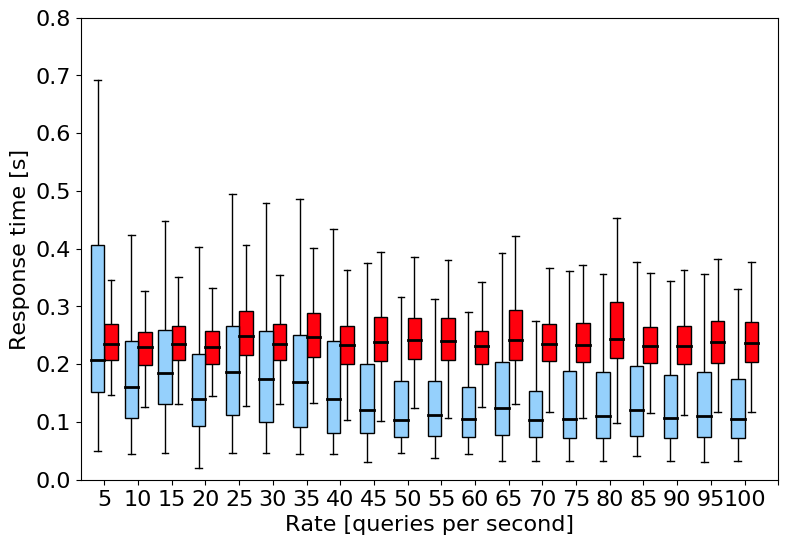}
    \caption{Throughput for 20 clients}
    \label{fig:throughput_20}
  \end{subfigure}
  \begin{subfigure}{0.32\textwidth}
    \centering
    \includegraphics[width=\textwidth]{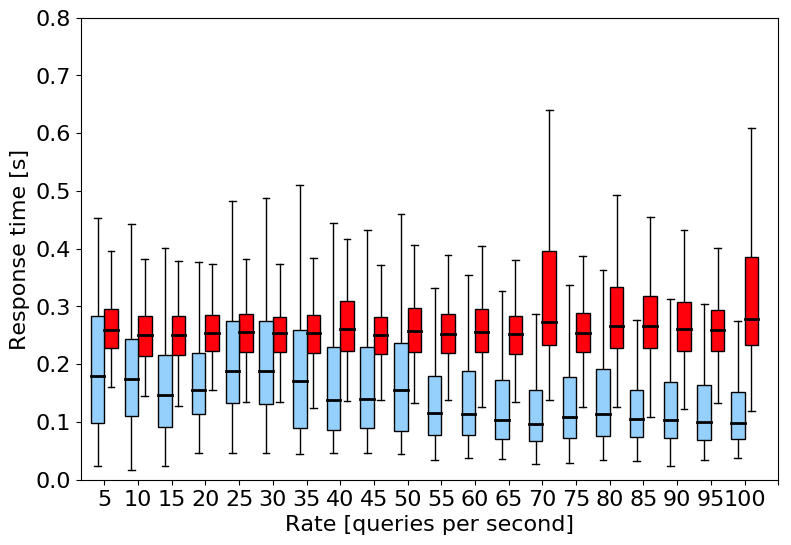}
    \caption{Throughput for 25 clients}
    \label{fig:throughput_25}
  \end{subfigure}
  \caption{Throughput comparison of \acron (red) and Unbound (blue)}
  \Description{Throughput comparison of \acron (red) and Unbound (blue)}
  \label{fig:throughput_eval}
\end{figure*}

\subsubsection{Throughput evaluation}\label{sec:throughput_eval}
The objective of throughput evaluation is to measure the rate at which the \RR 
can sustainably respond to queries. \acron's throughput should be close to that of Unbound to satisfy 
requirement R4.

\textbf{Experiment setup.} The client and \RR were run on different machines, so that the \RR could use all available 
resources of a single machine. This is representative of a local \RR running in a small network (e.g., a coffee 
shop WiFi network). We conducted this experiment using the same two \RR-s as in the latency experiment. 
Stubby was configured to reuse TLS connections.  To simulate a small to medium-scale network, we varied the 
number of concurrent clients between 1 and 25 and adjusted the query arrival rate from 5 to 100 queries per second.
Query rates were uniformly distributed among the clients, e.g., for an overall rate of 100 queries per second with 
10 clients, each client sends 10 queries per second.  To eliminate any variability in resolving the query, all queries 
were for the domain \texttt{google.com}. We maintained constant query rate for one minute.
Caching mechanisms of both \acron and Unbound were disabled.

\changed{

\begin{figure*}[t!]
  \centering
  \begin{subfigure}{0.32\textwidth}
    \centering
    \includegraphics[width=\textwidth]{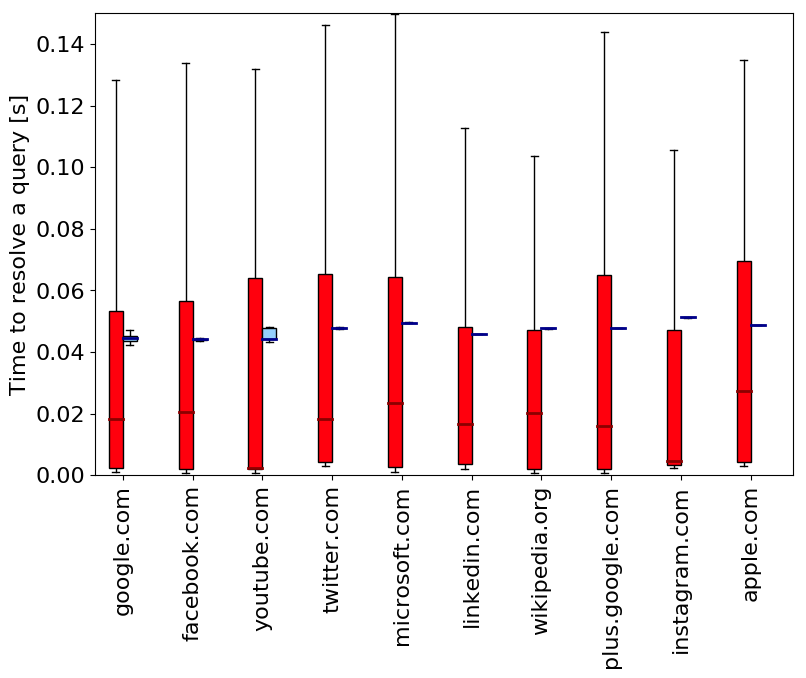}
    \caption{10 domains in cache}
    \label{fig:latency_cache_1}
  \end{subfigure}
  \begin{subfigure}{0.32\textwidth}
    \centering
    \includegraphics[width=\textwidth]{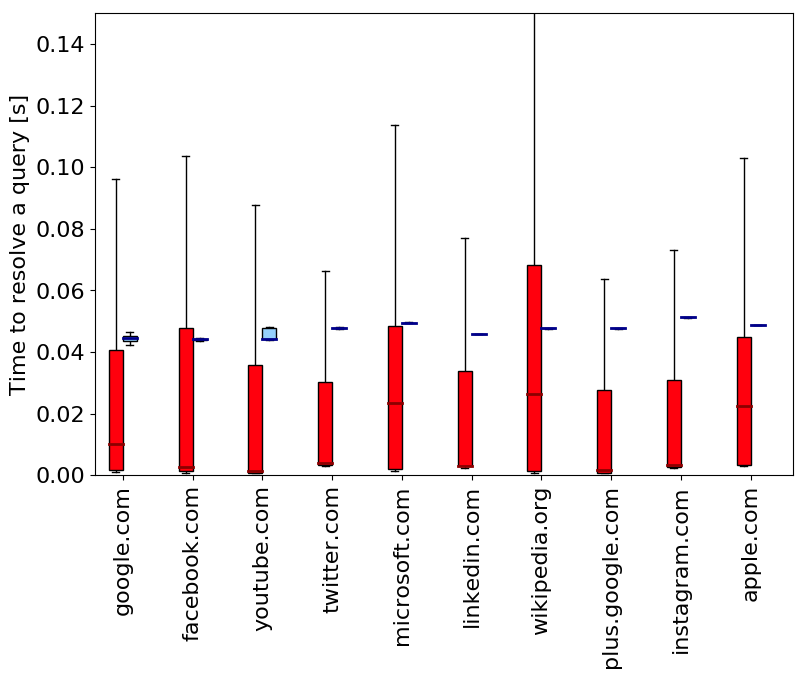}
    \caption{100 domains in cache}
    \label{fig:latency_cache_2}
  \end{subfigure}
  \begin{subfigure}{0.32\textwidth}
    \centering
    \includegraphics[width=\textwidth]{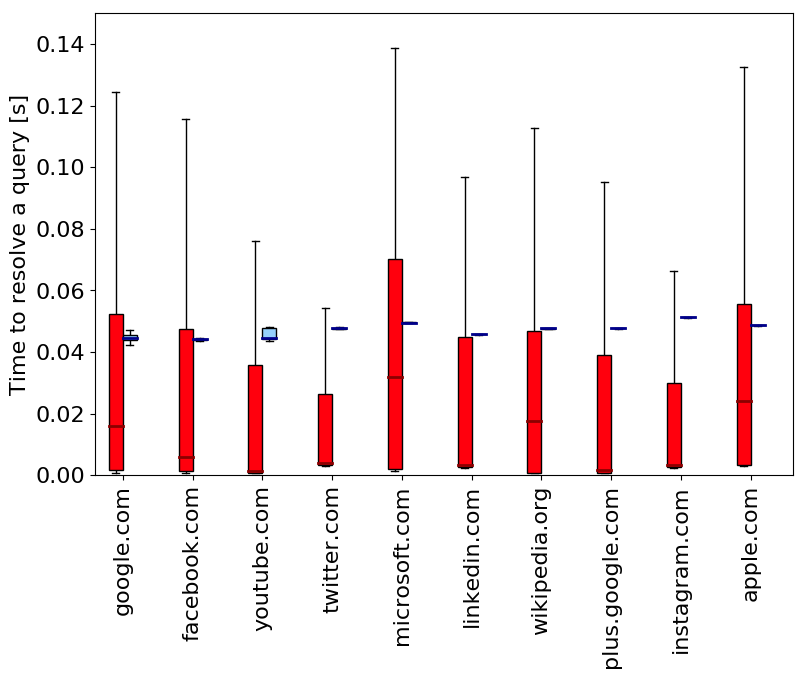}
    \caption{1000 domains in cache}
    \label{fig:latency_cache_3}
  \end{subfigure}
  \caption{Latency comparison of \acron (red) and Unbound (blue) with different number of domains in cache}
  \Description{Latency comparison of \acron (red) and Unbound (blue) with cache}
  \label{fig:latency_cache_eval}
\end{figure*}

\textbf{Results and observations.} Results of throughput experiments are shown in Figure~\ref{fig:throughput_eval}.
Each graph corresponds to a different number of clients.
Horizontal axis shows different query rates and vertical axis shows the range of response latencies for each query rate.
Measurement are plotted using the same box plot arrangement as in latency evaluation.
Blue boxes show results for Unbound and red boxes -- for \acron. 

If queries arrive faster than \RR can process them, they start clogging the queue, and the latency of each successive 
response increases as the queue grows. In this case, the average latency continues to increase indefinitely until 
queries begin to timeout or \RR runs out of memory. We view this rate of query arrival as \emph{unsustainable}.
On the other hand, if \RR can sustain the rate of query arrival, average response latency would remain roughly 
constant irrespective of how long \RR runs. For this experiment, we define a \emph{sustainable} rate of query 
arrival as the one for which the average response latency is constant over time, and below one second -- 
well below a typical DNS client timeout. Figure~\ref{fig:throughput_eval} only shows cases where query arrival rate 
is sustainable for the respective \RR. In other words, the presence of a box in Figure~\ref{fig:throughput_eval} 
shows that the \RR can achieve that level of throughput.

Surprisingly, we observed that Unbound cannot handle query rates exceeding 10 queries per second 
per client, i.e., its maximum sustainable rate was $10n$ queries per second distributed among $n$ clients.
This is because Unbound's design only uses one query processing thread per client.
In contrast, \acron handled more than 100 queries per second in all cases because its design uses a 
separate pool of QueryHandle threads.

Overall, Figure~\ref{fig:throughput_eval} confirms that our proof-of-concept implementation achieves at 
least the same throughput as Unbound across the range of clients and query arrival rates, and can 
achieve higher throughput when the number of clients is low. Although Unbound again achieves slightly 
lower latency, this is consistent with our latency measurements in Section~\ref{sec:latency_eval} and 
is likely due to the fact that Unbound is an optimized production-grade \RR.
}

\changed{
\subsubsection{Caching evaluation.}\label{sec:cache_eval}
We evaluated performance of both resolvers with caching enabled; Unbound with its default caching behavior, 
and \acron with our simple proof-of-concept cache.

\textbf{Experiment setup.} The experimental setup is similar to that of the latency evaluation described earlier.
We pre-populated resolvers' caches with varying numbers of domains and measured response latency for a 
representative set of 10 popular domains.

\textbf{Results and observations.} As shown in Figure~\ref{fig:latency_cache_eval}, Unbound serves responses 
from cache with a consistent latency irrespective of the number of entries in the cache.
Although \acron achieves lower average latencies when the cache is relatively empty, it has higher variability than Unbound.
This is probably due to the combination of our unoptimized caching implementation and latency of accessing enclave memory.
Nevertheless, Figure~\ref{fig:latency_cache_eval} shows that -- even with the memory limitations of current hardware enclaves -- 
\acron can still benefit from caching a small number of domains.

}

\section{Discussion}\label{sec:discussion}
\subsection{Information Revealed by IP Addresses}\label{sec:ppans}
Even if the connections between the client, \RR, and NS-s are encrypted using TLS, some information is still leaked.
The most prominent and obvious is source/destination IP addresses. 
The network adversary described in Section~\ref{sec:adv_model} can combine these cleartext IP addresses 
with packet timing information in order to correlate packets sent from client to \RR with subsequent 
packets sent from \RR to NS.

Armed with this information, the adversary can narrow down the client's domain name query to 
one of the records that could be served by that specific \ANS. Assuming the \ANS can serve $R$ 
domain names, the adversary has a $1/R$ probability of guessing which domain name the user queried.
When $R>1$, we call this a \textit{privacy-preserving \ANS}. This prompts two questions: 1) what percentage of
domains can be answered by a privacy-preserving \ANS; and 2) what is the typical size of anonymity 
set ($R$) provided by a privacy-preserving \ANS?

\begin{figure}[t!]
  \centering
  \includegraphics[width=0.49\textwidth]{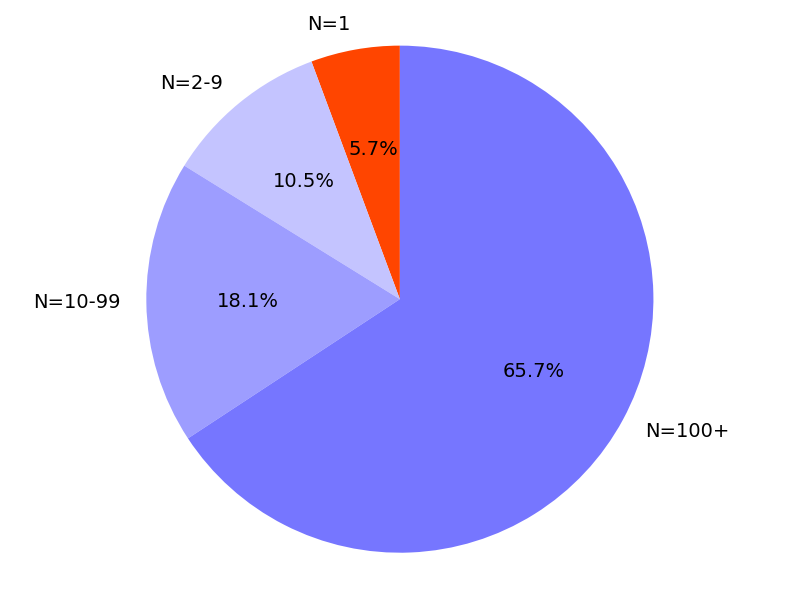}
  \caption{Percentage of Majestic Million domains answered by an \ANS with at least $N$ records}
  \Description{Percentage of Majestic Million domains answered by an \ANS with at least $N$ records}
  \label{fig:ans_records}
\end{figure}

To answer these questions, we designed a scheme to collect records stored in various \ANS-s. 
We sent DNS queries for 1,000,000 domains from the Majestic Million domain list~\cite{MajesticMillion}, and gathered information about \ANS-s that can possibly 
provide the answer for each. By collecting data on possible \ANS-s, we can map domain names to each 
\ANS, and thus estimate the number of records held by each \ANS. Following the \emph{Guidelines for 
Internet Measurement Activities}~\cite{Guidelines}, we limited our querying rate, in order to avoid placing 
undue load on any servers.
As shown in Figure~\ref{fig:ans_records}, only 5.7\% of domains we queried were served by non-privacy-preserving 
\ANS-s, i.e., those that hold only one record). Examples of domain names served from non-privacy-preserving \ANS-s included: 
\texttt{tinyurl.com}\footnote{Since \texttt{tinyurl.com} is a URL shortening service, this is actually still privacy-preserving 
because the adversary can not learn which short URL was queried.}, \texttt{bing.com}, \texttt{nginx.org}, \texttt{news.bbc.co.uk}, 
and \texttt{cloudflare.com}.
On the other hand, 9 out of 10 queries were served by a privacy-preserving \ANS, and 65.7\% by \ANS-s that hold over 100 records. 

\changed{
These results are still approximations.
Since we do not have data for domains outside the Majestic Million list, we cannot make claims about whether 
these would be served by a privacy-preserving \ANS.
We hypothesize that the vast majority of \ANS-s would be privacy-preserving for the simple reason that it is 
more economical to amortize the \ANS's running costs over multiple domains.
On the other hand, we can be certain that our results for the Majestic Million are a strict lower bound on the 
level of privacy because the \ANS-s from which these are served could also be serving other domains outside of our list.
It would be possible to arrive at a more accurate estimate by analyzing zone files of all (or at least most) \ANS-s.
However, virtually all \ANS-s disable the interface to download zone files because this could be used to mount DoS attacks.
Therefore, this type of analysis would have to be performed by an organization with privileged access to all \ANS-s' zone files.
}

\subsection{Caching \& Timing attacks}\label{sec:caching}
Introducing a cache into an \RR would allow the adversary to launch timing attacks and 
help guess the domain name queried by the end-user. We consider two types of timing attacks:
\begin{itemize}

  \item \textbf{Measuring time between query and response.} This is the simplest attack, whereby the 
  adversary monitors the network between client and \RR, and records the time for the \RR to respond 
  to the client. If the response time is shorter (compared to other queries), it likely has been served from 
  a cache. This attack can be launched by both adversary types described in Section~\ref{sec:adv_model}. 
  One obvious countermeasure is to artificially delay the response to match 
  the latency of NS-served responses.
  \item \textbf{Correlating client and \RR requests.} To counter the above countermeasure the adversary 
  may attempt to correlate DNS requests sent from client to \RR 
  with those sent from \RR to NS-s e.g., using the times at which the packets were sent. 
  If successful, the adversary can distinguish requests that involve contacting an NS from those that were served from the cache. This attack can be also launched by 
  a malicious \RR or a network sniffer. One way to counter this is to always send a query to one NS (although not necessarily the correct NS).
  This diminishes the benefits of caching, but still reduces the number of NS queries since the real answer may have required more than one NS query.  
  Another way is to batch and randomize the order of requests to the NS, creating a type of DNS mix network.
  
\end{itemize}

For both of these types of attacks, the information leakage depends on 
whether the adversary is passive or active. A passive adversary can (at most) guess the domain name. 
If the caching strategy is Most Recently Used (MRU), the domain name must be one of the 
popular ones. The active adversary can generate its own DNS queries for a wide range of domain names 
and keep a list of those that result in cache hits, thus improving the chances of inferring the user's query target.

\section{Related Work}\label{sec:related_work}
There has been much prior work aiming to protect the privacy of DNS
queries~\cite{AnonymousDNS,PIRDNS,AnalysisPDNS,PrivacyLeakDNS,PPDNS,PBP,ODNS}. 
For example, Lu et al.~\cite{PrivacyLeakDNS} proposed a privacy-preserving DNS that uses 
distributed hash tables, different naming schemes, and methods from computational private information retrieval. 
Federrath et al.~\cite{PPDNS} introduced a dedicated DNS Anonymity Service to protect the DNS queries
using an architecture that distributes the top domains by broadcast and uses low-latency mixes for requesting the
remaining domains. These schemes all assume that all parties involved do not act maliciously.

There have also been some activities in the Internet standards community that focused on DNS security and privacy. 
DNS Security Extensions (DNSSEC)~\cite{DNSSEC} provides data origin authentication and integrity 
via public key cryptography. However, it does not offer privacy. Bortzmeyer~\cite{MiniQuery} proposed a scheme 
Also, though not Internet standards, several protocols have been proposed to encrypt and authenticates DNS 
packets between the client and the \RR (DNSCrypt~\cite{DNSCrypt}) and \RR and NS-s (DNSCurve~\cite{DNSCurve}). 
Moreover, the original DNS-over-TLS paper has been converted into a draft Internet standard~\cite{DoTRFC}. 
All these methods assume that the \RR operator is trusted and does not attempt to learn anything from the DNS queries.

Furthermore, there has been some research on establishing trust through TEEs to protect confidentiality and integrity of 
network functions. Specifically, SGX has been used to protect network functions, especially middle-boxes. For example, 
Endbox~\cite{EndBox} aims to distribute middle-boxes to client edges: clients connect through VPN to ensure 
confidentiality of their traffic while remaining maintainable. LightBox~\cite{LightBox} is another middle-box that runs in 
an enclave; its goal is to protect the client's traffic from the third-party middle-box service provider while maintaining 
adequate performance. Finally, ShieldBox~\cite{ShieldBox} aims to protect confidential network traffic that flows through 
untrusted commodity servers and provides a generic interface for easy deployability. These efforts focus on 
protecting confidential data that flows in the network, and do not target DNS queries.

\section{Conclusion \& Future Work}\label{sec:conclusion}
This paper proposed \acron, a novel DNS \RR design that operates within a TEE to protect privacy of DNS 
queries, even from a malicious \RR operator.  In terms of query throughput, our unoptimized proof-of-concept 
implementation matches the throughput of Unbound, a state-of-the-art DNS-over-TLS recursive resolver, 
while incurring an acceptable increase in latency (due to the use of a TEE). 
In order to quantify the potential for privacy leakage through traffic analysis, we performed an 
Internet measurement study which showed that 94.7\% of the top 1,000,000 domain names can 
be served from a \emph{privacy-preserving} ANS that serves at least two distinct domain names, 
and 65.7\% from an ANS that serves 100+ domain names. 
As future work, we plan to port the Unbound \RR to Intel SGX and conduct a performance comparison 
with \acron, as well as to explore methods for improving \acron's performance using caching while maintaining client privacy.

\begin{acks}
We thank Geonhee Cho for the initial data collection for the privacy-preserving \ANS analysis in Section~\ref{sec:ppans}.
We are also grateful to the paper's shepherd, Roberto Perdisci, and ACSAC'19 anonymous reviewers for their valuable comments.
First and third authors were supported in part by NSF Award Number:1840197,
titled: "CICI: SSC: Horizon: Secure Large-Scale Scientific Cloud Computing".
The first author was also supported by The Nakajima Foundation.
The second author was supported by a US-UK Fulbright Cyber Security Scholar Award.
\end{acks}

\balance
\bibliographystyle{ACM-Reference-Format}
\bibliography{dots}


\begin{thebibliography}{39}


\ifx \showCODEN    \undefined \def \showCODEN     #1{\unskip}     \fi
\ifx \showDOI      \undefined \def \showDOI       #1{#1}\fi
\ifx \showISBNx    \undefined \def \showISBNx     #1{\unskip}     \fi
\ifx \showISBNxiii \undefined \def \showISBNxiii  #1{\unskip}     \fi
\ifx \showISSN     \undefined \def \showISSN      #1{\unskip}     \fi
\ifx \showLCCN     \undefined \def \showLCCN      #1{\unskip}     \fi
\ifx \shownote     \undefined \def \shownote      #1{#1}          \fi
\ifx \showarticletitle \undefined \def \showarticletitle #1{#1}   \fi
\ifx \showURL      \undefined \def \showURL       {\relax}        \fi
\providecommand\bibfield[2]{#2}
\providecommand\bibinfo[2]{#2}
\providecommand\natexlab[1]{#1}
\providecommand\showeprint[2][]{arXiv:#2}

\bibitem[\protect\citeauthoryear{}{}{2009}]%
        {DNSCurve}
\bibfield{author}{\bibinfo{person}{}.} \bibinfo{year}{2009}\natexlab{}.
\newblock \bibinfo{title}{{Introduction to DNSCurve}}.
\newblock
\newblock
\urldef\tempurl%
\url{https://dnscurve.org/index.html}
\showURL{%
\tempurl}
\newblock
\shownote{[Online] Accessed: 2019-05-29.}


\bibitem[\protect\citeauthoryear{Alder, Asokan, Kurnikov, Paverd, and
  Steiner}{Alder et~al\mbox{.}}{2019}]%
        {alder2019sfaas}
\bibfield{author}{\bibinfo{person}{Fritz Alder}, \bibinfo{person}{N Asokan},
  \bibinfo{person}{Arseny Kurnikov}, \bibinfo{person}{Andrew Paverd}, {and}
  \bibinfo{person}{Michael Steiner}.} \bibinfo{year}{2019}\natexlab{}.
\newblock \showarticletitle{S-FaaS: Trustworthy and Accountable
  Function-as-a-Service using Intel SGX}. In \bibinfo{booktitle}{\emph{ACM
  Cloud Computing Security Workshop}} \emph{(\bibinfo{series}{CCSW '19})}.
\newblock


\bibitem[\protect\citeauthoryear{Arends, Austein, Larson, Massey, and
  Rose}{Arends et~al\mbox{.}}{2005}]%
        {DNSSEC}
\bibfield{author}{\bibinfo{person}{R. Arends}, \bibinfo{person}{R. Austein},
  \bibinfo{person}{M. Larson}, \bibinfo{person}{D. Massey}, {and}
  \bibinfo{person}{S. Rose}.} \bibinfo{year}{2005}\natexlab{}.
\newblock \bibinfo{booktitle}{\emph{{DNS Security Introduction and
  Requirements}}}.
\newblock \bibinfo{type}{{T}echnical {R}eport}.
\newblock
\urldef\tempurl%
\url{https://doi.org/10.17487/rfc4033}
\showDOI{\tempurl}


\bibitem[\protect\citeauthoryear{ARM}{ARM}{2009}]%
        {TZ}
\bibfield{author}{\bibinfo{person}{ARM}.} \bibinfo{year}{2009}\natexlab{}.
\newblock \bibinfo{title}{{ARM Security Technology - Building a Secure System
  using TrustZone Technology}}.
\newblock
\newblock
\urldef\tempurl%
\url{http://infocenter.arm.com/help/index.jsp?topic=/com.arm.doc.prd29-genc-009492c/index.html}
\showURL{%
\tempurl}
\newblock
\shownote{[Online] Accessed: 2019-05-29.}


\bibitem[\protect\citeauthoryear{Bortzmeyer}{Bortzmeyer}{2016}]%
        {MiniQuery}
\bibfield{author}{\bibinfo{person}{S. Bortzmeyer}.}
  \bibinfo{year}{2016}\natexlab{}.
\newblock \bibinfo{booktitle}{\emph{{DNS Query Name Minimisation to Improve
  Privacy}}}.
\newblock \bibinfo{type}{{T}echnical {R}eport}.
\newblock
\urldef\tempurl%
\url{https://doi.org/10.17487/RFC7816}
\showDOI{\tempurl}


\bibitem[\protect\citeauthoryear{Bortzmeyer}{Bortzmeyer}{2018}]%
        {dns-res-to-auth}
\bibfield{author}{\bibinfo{person}{S Bortzmeyer}.}
  \bibinfo{year}{2018}\natexlab{}.
\newblock \bibinfo{title}{{Encryption and authentication of the DNS
  resolver-to-authoritative communication}}.
\newblock
\newblock
\urldef\tempurl%
\url{https://tools.ietf.org/html/draft-bortzmeyer-dprive-resolver-to-auth-01}
\showURL{%
\tempurl}


\bibitem[\protect\citeauthoryear{Castillo-Perez and
  Garcia-Alfaro}{Castillo-Perez and Garcia-Alfaro}{2008}]%
        {AnonymousDNS}
\bibfield{author}{\bibinfo{person}{Sergio Castillo-Perez} {and}
  \bibinfo{person}{Joaquin Garcia-Alfaro}.} \bibinfo{year}{2008}\natexlab{}.
\newblock \showarticletitle{{Anonymous Resolution of DNS Queries}}.
\newblock \bibinfo{publisher}{Springer, Berlin, Heidelberg},
  \bibinfo{pages}{987--1000}.
\newblock
\urldef\tempurl%
\url{https://doi.org/10.1007/978-3-540-88873-4_5}
\showDOI{\tempurl}


\bibitem[\protect\citeauthoryear{Cerf}{Cerf}{1991}]%
        {Guidelines}
\bibfield{author}{\bibinfo{person}{V.G. Cerf}.}
  \bibinfo{year}{1991}\natexlab{}.
\newblock \bibinfo{booktitle}{\emph{{Guidelines for Internet Measurement
  Activities}}}.
\newblock \bibinfo{type}{{T}echnical {R}eport}.
\newblock
\urldef\tempurl%
\url{https://doi.org/10.17487/rfc1262}
\showDOI{\tempurl}


\bibitem[\protect\citeauthoryear{Cloudflare}{Cloudflare}{ }]%
        {Cloudflare}
\bibfield{author}{\bibinfo{person}{Cloudflare}.} \bibinfo{year}{-}\natexlab{}.
\newblock \bibinfo{title}{{DNS over TLS - Cloudflare Resolver}}.
\newblock
\newblock
\urldef\tempurl%
\url{https://1.1.1.1/dns/}
\showURL{%
\tempurl}
\newblock
\shownote{[Online] Accessed: 2019-05-29.}


\bibitem[\protect\citeauthoryear{Costa, Esswood, Ohrimenko, Schuster, and
  Wagh}{Costa et~al\mbox{.}}{2017}]%
        {costa2017pyramid}
\bibfield{author}{\bibinfo{person}{Manuel Costa}, \bibinfo{person}{Lawrence
  Esswood}, \bibinfo{person}{Olga Ohrimenko}, \bibinfo{person}{Felix Schuster},
  {and} \bibinfo{person}{Sameer Wagh}.} \bibinfo{year}{2017}\natexlab{}.
\newblock \showarticletitle{The pyramid scheme: Oblivious RAM for trusted
  processors}.
\newblock \bibinfo{journal}{\emph{arXiv preprint arXiv:1712.07882}}
  (\bibinfo{year}{2017}).
\newblock


\bibitem[\protect\citeauthoryear{Costan, Lebedev, and Devadas}{Costan
  et~al\mbox{.}}{2016}]%
        {Sanctum}
\bibfield{author}{\bibinfo{person}{Victor Costan}, \bibinfo{person}{Ilia
  Lebedev}, {and} \bibinfo{person}{Srinivas Devadas}.}
  \bibinfo{year}{2016}\natexlab{}.
\newblock \bibinfo{title}{{Sanctum: Minimal Hardware Extensions for Strong
  Software Isolation}}.
\newblock , \bibinfo{numpages}{857--874}~pages.
\newblock
\urldef\tempurl%
\url{https://www.usenix.org/conference/usenixsecurity16/technical-sessions/presentation/costan}
\showURL{%
\tempurl}


\bibitem[\protect\citeauthoryear{cs.nic}{cs.nic}{ }]%
        {KnotResolver}
\bibfield{author}{\bibinfo{person}{cs.nic}.} \bibinfo{year}{-}\natexlab{}.
\newblock \bibinfo{title}{{Knot Resolver}}.
\newblock
\newblock
\urldef\tempurl%
\url{https://www.knot-resolver.cz/}
\showURL{%
\tempurl}
\newblock
\shownote{[Online] Accessed: 2019-05-29.}


\bibitem[\protect\citeauthoryear{Dierks and Rescorla}{Dierks and
  Rescorla}{2008}]%
        {TLS}
\bibfield{author}{\bibinfo{person}{T. Dierks} {and} \bibinfo{person}{E.
  Rescorla}.} \bibinfo{year}{2008}\natexlab{}.
\newblock \bibinfo{booktitle}{\emph{{The Transport Layer Security (TLS)
  Protocol Version 1.2}}}.
\newblock \bibinfo{type}{{T}echnical {R}eport}.
\newblock
\urldef\tempurl%
\url{https://doi.org/10.17487/rfc5246}
\showDOI{\tempurl}


\bibitem[\protect\citeauthoryear{Duan, Wang, Yuan, Zhou, Wang, and Ren}{Duan
  et~al\mbox{.}}{2017}]%
        {LightBox}
\bibfield{author}{\bibinfo{person}{Huayi Duan}, \bibinfo{person}{Cong Wang},
  \bibinfo{person}{Xingliang Yuan}, \bibinfo{person}{Yajin Zhou},
  \bibinfo{person}{Qian Wang}, {and} \bibinfo{person}{Kui Ren}.}
  \bibinfo{year}{2017}\natexlab{}.
\newblock \showarticletitle{{LightBox: Full-stack Protected Stateful Middlebox
  at Lightning Speed}}.
\newblock  (\bibinfo{date}{Jun} \bibinfo{year}{2017}).
\newblock
\showeprint[arxiv]{1706.06261}
\urldef\tempurl%
\url{http://arxiv.org/abs/1706.06261}
\showURL{%
\tempurl}


\bibitem[\protect\citeauthoryear{Edmundson, Schmitt, and Feamster}{Edmundson
  et~al\mbox{.}}{2018}]%
        {ODNS}
\bibfield{author}{\bibinfo{person}{Annie Edmundson}, \bibinfo{person}{Paul
  Schmitt}, {and} \bibinfo{person}{Nick Feamster}.}
  \bibinfo{year}{2018}\natexlab{}.
\newblock \bibinfo{title}{{ODNS: Oblivious DNS}}.
\newblock
\newblock
\urldef\tempurl%
\url{https://odns.cs.princeton.edu/}
\showURL{%
\tempurl}
\newblock
\shownote{[Online] Accessed: 2019-05-29.}


\bibitem[\protect\citeauthoryear{Federrath, Fuchs, Herrmann, and
  Piosecny}{Federrath et~al\mbox{.}}{2011}]%
        {PPDNS}
\bibfield{author}{\bibinfo{person}{Hannes Federrath},
  \bibinfo{person}{Karl-Peter Fuchs}, \bibinfo{person}{Dominik Herrmann}, {and}
  \bibinfo{person}{Christopher Piosecny}.} \bibinfo{year}{2011}\natexlab{}.
\newblock \showarticletitle{{Privacy-Preserving DNS: Analysis of Broadcast,
  Range Queries and Mix-Based Protection Methods}}.
\newblock \bibinfo{publisher}{Springer, Berlin, Heidelberg},
  \bibinfo{pages}{665--683}.
\newblock
\urldef\tempurl%
\url{https://doi.org/10.1007/978-3-642-23822-2_36}
\showDOI{\tempurl}


\bibitem[\protect\citeauthoryear{Goltzsche, Rusch, Nieke, Vaucher, Weichbrodt,
  Schiavoni, Aublin, Cosa, Fetzer, Felber, Pietzuch, and Kapitza}{Goltzsche
  et~al\mbox{.}}{2018}]%
        {EndBox}
\bibfield{author}{\bibinfo{person}{David Goltzsche}, \bibinfo{person}{Signe
  Rusch}, \bibinfo{person}{Manuel Nieke}, \bibinfo{person}{Sebastien Vaucher},
  \bibinfo{person}{Nico Weichbrodt}, \bibinfo{person}{Valerio Schiavoni},
  \bibinfo{person}{Pierre-Louis Aublin}, \bibinfo{person}{Paolo Cosa},
  \bibinfo{person}{Christof Fetzer}, \bibinfo{person}{Pascal Felber},
  \bibinfo{person}{Peter Pietzuch}, {and} \bibinfo{person}{Rudiger Kapitza}.}
  \bibinfo{year}{2018}\natexlab{}.
\newblock \showarticletitle{{EndBox: Scalable Middlebox Functions Using
  Client-Side Trusted Execution}}. In \bibinfo{booktitle}{\emph{2018 48th
  Annual IEEE/IFIP International Conference on Dependable Systems and
  Networks}} \emph{(\bibinfo{series}{DSN '18})}. \bibinfo{publisher}{IEEE},
  \bibinfo{pages}{386--397}.
\newblock
\showISBNx{978-1-5386-5596-2}
\urldef\tempurl%
\url{https://doi.org/10.1109/DSN.2018.00048}
\showDOI{\tempurl}


\bibitem[\protect\citeauthoryear{Google}{Google}{2018}]%
        {AndroidP}
\bibfield{author}{\bibinfo{person}{Google}.} \bibinfo{year}{2018}\natexlab{}.
\newblock \bibinfo{title}{{DNS over TLS support in Android P Developer
  Preview}}.
\newblock
\newblock
\urldef\tempurl%
\url{https://security.googleblog.com/2018/04/dns-over-tls-support-in-android-p.html}
\showURL{%
\tempurl}
\newblock
\shownote{[Online] Accessed: 2019-05-29.}


\bibitem[\protect\citeauthoryear{Hu, Zhu, Heidemann, Mankin, Wessels, and
  Hoffman}{Hu et~al\mbox{.}}{2016}]%
        {DoTRFC}
\bibfield{author}{\bibinfo{person}{Zi Hu}, \bibinfo{person}{Liang Zhu},
  \bibinfo{person}{John Heidemann}, \bibinfo{person}{Allison Mankin},
  \bibinfo{person}{Duane Wessels}, {and} \bibinfo{person}{P Hoffman}.}
  \bibinfo{year}{2016}\natexlab{}.
\newblock \bibinfo{title}{{Specification for DNS over Transport Layer Security
  (TLS)}}.
\newblock
\newblock
\urldef\tempurl%
\url{https://doi.org/10.17487/RFC7858}
\showDOI{\tempurl}


\bibitem[\protect\citeauthoryear{Knauth, Steiner, Chakrabarti, Lei, Xing, and
  Vij}{Knauth et~al\mbox{.}}{2018}]%
        {SGXRATLS}
\bibfield{author}{\bibinfo{person}{Thomas Knauth}, \bibinfo{person}{Michael
  Steiner}, \bibinfo{person}{Somnath Chakrabarti}, \bibinfo{person}{Li Lei},
  \bibinfo{person}{Cedric Xing}, {and} \bibinfo{person}{Mona Vij}.}
  \bibinfo{year}{2018}\natexlab{}.
\newblock \showarticletitle{{Integrating Remote Attestation with Transport
  Layer Security}}.
\newblock  (\bibinfo{date}{Jan} \bibinfo{year}{2018}).
\newblock
\showeprint[arxiv]{1801.05863}
\urldef\tempurl%
\url{http://arxiv.org/abs/1801.05863}
\showURL{%
\tempurl}


\bibitem[\protect\citeauthoryear{Lab}{Lab}{2019}]%
        {PDoTcode}
\bibfield{author}{\bibinfo{person}{SPROUT Lab}.}
  \bibinfo{year}{2019}\natexlab{}.
\newblock \bibinfo{title}{{PDoT Source Code}}.
\newblock
\newblock
\urldef\tempurl%
\url{https://github.com/sprout-uci/PDoT}
\showURL{%
\tempurl}


\bibitem[\protect\citeauthoryear{Labs}{Labs}{ a}]%
        {Stubby}
\bibfield{author}{\bibinfo{person}{NLnet Labs}.} \bibinfo{year}{-}\natexlab{a}.
\newblock \bibinfo{title}{{Stubby}}.
\newblock
\newblock
\urldef\tempurl%
\url{https://dnsprivacy.org/wiki/display/DP/DNS+Privacy+Daemon+-+Stubby}
\showURL{%
\tempurl}
\newblock
\shownote{[Online] Accessed: 2019-05-29.}


\bibitem[\protect\citeauthoryear{Labs}{Labs}{ b}]%
        {Unbound}
\bibfield{author}{\bibinfo{person}{NLnet Labs}.} \bibinfo{year}{-}\natexlab{b}.
\newblock \bibinfo{title}{{Unbound}}.
\newblock
\newblock
\urldef\tempurl%
\url{https://nlnetlabs.nl/projects/unbound/about/}
\showURL{%
\tempurl}
\newblock
\shownote{[Online] Accessed: 2019-05-29.}


\bibitem[\protect\citeauthoryear{Liu, Yarom, Ge, Heiser, and Lee}{Liu
  et~al\mbox{.}}{2015}]%
        {CacheSideChannel}
\bibfield{author}{\bibinfo{person}{Fangfei Liu}, \bibinfo{person}{Yuval Yarom},
  \bibinfo{person}{Qian Ge}, \bibinfo{person}{Gernot Heiser}, {and}
  \bibinfo{person}{Ruby~B. Lee}.} \bibinfo{year}{2015}\natexlab{}.
\newblock \showarticletitle{{Last-Level Cache Side-Channel Attacks are
  Practical}}. In \bibinfo{booktitle}{\emph{2015 IEEE Symposium on Security and
  Privacy}}. \bibinfo{publisher}{IEEE}, \bibinfo{pages}{605--622}.
\newblock
\showISBNx{978-1-4673-6949-7}
\urldef\tempurl%
\url{https://doi.org/10.1109/SP.2015.43}
\showDOI{\tempurl}


\bibitem[\protect\citeauthoryear{Lu and Tsudik}{Lu and Tsudik}{2010}]%
        {PrivacyLeakDNS}
\bibfield{author}{\bibinfo{person}{Y. Lu} {and} \bibinfo{person}{G. Tsudik}.}
  \bibinfo{year}{2010}\natexlab{}.
\newblock \showarticletitle{{Towards Plugging Privacy Leaks in the Domain Name
  System}}. In \bibinfo{booktitle}{\emph{2010 IEEE Tenth International
  Conference on Peer-to-Peer Computing}} \emph{(\bibinfo{series}{P2P})}.
  \bibinfo{publisher}{IEEE}, \bibinfo{pages}{1--10}.
\newblock
\showISBNx{978-1-4244-7140-9}
\urldef\tempurl%
\url{https://doi.org/10.1109/P2P.2010.5569976}
\showDOI{\tempurl}


\bibitem[\protect\citeauthoryear{Majestic}{Majestic}{2012}]%
        {MajesticMillion}
\bibfield{author}{\bibinfo{person}{Majestic}.} \bibinfo{year}{2012}\natexlab{}.
\newblock \bibinfo{title}{{Majestic Million}}.
\newblock
\newblock
\urldef\tempurl%
\url{https://blog.majestic.com/development/majestic-million-csv-daily/}
\showURL{%
\tempurl}


\bibitem[\protect\citeauthoryear{McKeen, Alexandrovich, Berenzon, Rozas, Shafi,
  Shanbhogue, and Savagaonkar}{McKeen et~al\mbox{.}}{2013}]%
        {SGX}
\bibfield{author}{\bibinfo{person}{Frank McKeen}, \bibinfo{person}{Ilya
  Alexandrovich}, \bibinfo{person}{Alex Berenzon}, \bibinfo{person}{Carlos~V.
  Rozas}, \bibinfo{person}{Hisham Shafi}, \bibinfo{person}{Vedvyas Shanbhogue},
  {and} \bibinfo{person}{Uday~R. Savagaonkar}.}
  \bibinfo{year}{2013}\natexlab{}.
\newblock \showarticletitle{{Innovative instructions and software model for
  isolated execution}}. In \bibinfo{booktitle}{\emph{Proceedings of the 2nd
  International Workshop on Hardware and Architectural Support for Security and
  Privacy}} \emph{(\bibinfo{series}{HASP '13})}. \bibinfo{publisher}{ACM
  Press}, \bibinfo{address}{New York, New York, USA}, \bibinfo{pages}{1--1}.
\newblock
\showISBNx{9781450321181}
\urldef\tempurl%
\url{https://doi.org/10.1145/2487726.2488368}
\showDOI{\tempurl}


\bibitem[\protect\citeauthoryear{Microsoft}{Microsoft}{2017}]%
        {AzureSGX}
\bibfield{author}{\bibinfo{person}{Microsoft}.}
  \bibinfo{year}{2017}\natexlab{}.
\newblock \bibinfo{title}{{Introducing Azure confidential computing}}.
\newblock
\newblock
\urldef\tempurl%
\url{https://azure.microsoft.com/en-us/blog/introducing-azure-confidential-computing/}
\showURL{%
\tempurl}
\newblock
\shownote{[Online] Accessed: 2019-05-29.}


\bibitem[\protect\citeauthoryear{Mockapetris}{Mockapetris}{1987}]%
        {DNS}
\bibfield{author}{\bibinfo{person}{P.V. Mockapetris}.}
  \bibinfo{year}{1987}\natexlab{}.
\newblock \bibinfo{booktitle}{\emph{{Domain names - implementation and
  specification}}}.
\newblock \bibinfo{type}{{T}echnical {R}eport}.
\newblock
\urldef\tempurl%
\url{https://doi.org/10.17487/rfc1035}
\showDOI{\tempurl}


\bibitem[\protect\citeauthoryear{Project}{Project}{ }]%
        {DNSCrypt}
\bibfield{author}{\bibinfo{person}{DNSCrypt Project}.}
  \bibinfo{year}{-}\natexlab{}.
\newblock \bibinfo{title}{{DNSCrypt}}.
\newblock
\newblock
\urldef\tempurl%
\url{https://dnscrypt.info/}
\showURL{%
\tempurl}
\newblock
\shownote{[Online] Accessed: 2019-05-29.}


\bibitem[\protect\citeauthoryear{Sasy, Gorbunov, and Fletcher}{Sasy
  et~al\mbox{.}}{2017}]%
        {sasy2017zerotrace}
\bibfield{author}{\bibinfo{person}{Sajin Sasy}, \bibinfo{person}{Sergey
  Gorbunov}, {and} \bibinfo{person}{Christopher~W Fletcher}.}
  \bibinfo{year}{2017}\natexlab{}.
\newblock \showarticletitle{ZeroTrace: Oblivious Memory Primitives from Intel
  SGX.}
\newblock \bibinfo{journal}{\emph{IACR Cryptology ePrint Archive}}
  \bibinfo{volume}{2017} (\bibinfo{year}{2017}), \bibinfo{pages}{549}.
\newblock


\bibitem[\protect\citeauthoryear{Shih, Lee, Kim, and Peinado}{Shih
  et~al\mbox{.}}{2017}]%
        {TSGX}
\bibfield{author}{\bibinfo{person}{Ming-Wei Shih}, \bibinfo{person}{Sangho
  Lee}, \bibinfo{person}{Taesoo Kim}, {and} \bibinfo{person}{Marcus Peinado}.}
  \bibinfo{year}{2017}\natexlab{}.
\newblock \showarticletitle{{T-SGX: Eradicating Controlled-Channel Attacks
  Against Enclave Programs}}. In \bibinfo{booktitle}{\emph{NDSS Symposium}}.
\newblock
\urldef\tempurl%
\url{https://www.ndss-symposium.org/ndss2017/ndss-2017-programme/t-sgx-eradicating-controlled-channel-attacks-against-enclave-programs/}
\showURL{%
\tempurl}


\bibitem[\protect\citeauthoryear{Shulman and Haya}{Shulman and Haya}{2014}]%
        {PBP}
\bibfield{author}{\bibinfo{person}{Haya Shulman} {and} \bibinfo{person}{Haya}.}
  \bibinfo{year}{2014}\natexlab{}.
\newblock \showarticletitle{{Pretty Bad Privacy: Pitfalls of DNS Encryption}}.
  In \bibinfo{booktitle}{\emph{Proceedings of the 13th Workshop on Privacy in
  the Electronic Society}} \emph{(\bibinfo{series}{WPES '14})}.
  \bibinfo{publisher}{ACM Press}, \bibinfo{address}{New York, New York, USA},
  \bibinfo{pages}{191--200}.
\newblock
\showISBNx{9781450331487}
\urldef\tempurl%
\url{https://doi.org/10.1145/2665943.2665959}
\showDOI{\tempurl}


\bibitem[\protect\citeauthoryear{Tamrakar, Liu, Paverd, Ekberg, Pinkas, and
  Asokan}{Tamrakar et~al\mbox{.}}{2017}]%
        {Tamrakar2017}
\bibfield{author}{\bibinfo{person}{Sandeep Tamrakar}, \bibinfo{person}{Jian
  Liu}, \bibinfo{person}{Andrew Paverd}, \bibinfo{person}{Jan-Erik Ekberg},
  \bibinfo{person}{Benny Pinkas}, {and} \bibinfo{person}{N. Asokan}.}
  \bibinfo{year}{2017}\natexlab{}.
\newblock \showarticletitle{The Circle Game: Scalable Private Membership Test
  Using Trusted Hardware}. In \bibinfo{booktitle}{\emph{Proceedings of the 2017
  ACM on Asia Conference on Computer and Communications Security}}
  \emph{(\bibinfo{series}{ASIA CCS '17})}.
\newblock
\showISBNx{978-1-4503-4944-4}
\urldef\tempurl%
\url{https://doi.org/10.1145/3052973.3053006}
\showDOI{\tempurl}


\bibitem[\protect\citeauthoryear{Trach, Krohmer, Gregor, Arnautov, Bhatotia,
  and Fetzer}{Trach et~al\mbox{.}}{2018}]%
        {ShieldBox}
\bibfield{author}{\bibinfo{person}{Bohdan Trach}, \bibinfo{person}{Alfred
  Krohmer}, \bibinfo{person}{Franz Gregor}, \bibinfo{person}{Sergei Arnautov},
  \bibinfo{person}{Pramod Bhatotia}, {and} \bibinfo{person}{Christof Fetzer}.}
  \bibinfo{year}{2018}\natexlab{}.
\newblock \showarticletitle{{ShieldBox: Secure Middleboxes using Shielded
  Execution}}. In \bibinfo{booktitle}{\emph{Proceedings of the Symposium on SDN
  Research}} \emph{(\bibinfo{series}{SOSR '18})}. \bibinfo{publisher}{ACM
  Press}, \bibinfo{address}{New York, New York, USA}, \bibinfo{pages}{1--14}.
\newblock
\showISBNx{9781450356640}
\urldef\tempurl%
\url{https://doi.org/10.1145/3185467.3185469}
\showDOI{\tempurl}


\bibitem[\protect\citeauthoryear{Xu, Cui, and Peinado}{Xu
  et~al\mbox{.}}{2015}]%
        {ControllChannelSideChannel}
\bibfield{author}{\bibinfo{person}{Yuanzhong Xu}, \bibinfo{person}{Weidong
  Cui}, {and} \bibinfo{person}{Marcus Peinado}.}
  \bibinfo{year}{2015}\natexlab{}.
\newblock \showarticletitle{{Controlled-Channel Attacks: Deterministic Side
  Channels for Untrusted Operating Systems}}. In \bibinfo{booktitle}{\emph{2015
  IEEE Symposium on Security and Privacy}}. \bibinfo{publisher}{IEEE},
  \bibinfo{pages}{640--656}.
\newblock
\showISBNx{978-1-4673-6949-7}
\urldef\tempurl%
\url{https://doi.org/10.1109/SP.2015.45}
\showDOI{\tempurl}


\bibitem[\protect\citeauthoryear{Zhao, Hori, and Sakurai}{Zhao
  et~al\mbox{.}}{2007a}]%
        {AnalysisPDNS}
\bibfield{author}{\bibinfo{person}{Fangming Zhao}, \bibinfo{person}{Yoshiaki
  Hori}, {and} \bibinfo{person}{Kouichi Sakurai}.}
  \bibinfo{year}{2007}\natexlab{a}.
\newblock \showarticletitle{{Analysis of Privacy Disclosure in DNS Query}}. In
  \bibinfo{booktitle}{\emph{2007 International Conference on Multimedia and
  Ubiquitous Engineering}} \emph{(\bibinfo{series}{MUE '07})}.
  \bibinfo{publisher}{IEEE}, \bibinfo{pages}{952--957}.
\newblock
\showISBNx{0-7695-2777-9}
\urldef\tempurl%
\url{https://doi.org/10.1109/MUE.2007.84}
\showDOI{\tempurl}


\bibitem[\protect\citeauthoryear{Zhao, Hori, and Sakurai}{Zhao
  et~al\mbox{.}}{2007b}]%
        {PIRDNS}
\bibfield{author}{\bibinfo{person}{Fangming Zhao}, \bibinfo{person}{Yoshiaki
  Hori}, {and} \bibinfo{person}{Kouichi Sakurai}.}
  \bibinfo{year}{2007}\natexlab{b}.
\newblock \showarticletitle{{Two-Servers PIR Based DNS Query Scheme with
  Privacy-Preserving}}. In \bibinfo{booktitle}{\emph{The 2007 International
  Conference on Intelligent Pervasive Computing}} \emph{(\bibinfo{series}{IPC
  '07})}. \bibinfo{publisher}{IEEE}, \bibinfo{pages}{299--302}.
\newblock
\showISBNx{978-0-7695-3006-2}
\urldef\tempurl%
\url{https://doi.org/10.1109/IPC.2007.27}
\showDOI{\tempurl}


\bibitem[\protect\citeauthoryear{Zhu, Hu, Heidemann, Wessels, Mankin, and
  Somaiya}{Zhu et~al\mbox{.}}{2015}]%
        {DoTOakland}
\bibfield{author}{\bibinfo{person}{Liang Zhu}, \bibinfo{person}{Zi Hu},
  \bibinfo{person}{John Heidemann}, \bibinfo{person}{Duane Wessels},
  \bibinfo{person}{Allison Mankin}, {and} \bibinfo{person}{Nikita Somaiya}.}
  \bibinfo{year}{2015}\natexlab{}.
\newblock \showarticletitle{{Connection-Oriented DNS to Improve Privacy and
  Security}}. In \bibinfo{booktitle}{\emph{2015 IEEE Symposium on Security and
  Privacy}}. \bibinfo{publisher}{IEEE}, \bibinfo{pages}{171--186}.
\newblock
\showISBNx{978-1-4673-6949-7}
\urldef\tempurl%
\url{https://doi.org/10.1109/SP.2015.18}
\showDOI{\tempurl}


\end{thebibliography}

\end{document}